\documentclass[twocolumn,letterpaper,headsepline,twoside,10pt,smallheadings]{scrartcl}

\typearea[8mm]{26}  

\usepackage{times}

\renewcommand{\rightmark}{GADGET: A code for collisionless and gasdynamical cosmological simulations}
\usepackage{graphics}
\usepackage{natbib}
\bibpunct[,]{(}{)}{;}{a}{}{,}

\newcommand{\be}{\begin{equation}}
\newcommand{\ee}{\end{equation}}
\newcommand{\bea}{\begin{eqnarray}}
\newcommand{\eea}{\end{eqnarray}}
\newcommand{\bc}{\begin{center}}
\newcommand{\ec}{\end{center}}
\newcommand{\ol}[1]{ {\overline{#1}}}
\newcommand{\lu}{\,h^{-1}{\rm kpc}}
\newcommand{\mlu}{\,h^{-1}{\rm Mpc}}
\newcommand{\Mlu}{\,h^{-1}{\rm Mpc}}
\renewcommand{\vec}[1]{ {\bmath #1} } 
\newcommand{\dd}{{\rm d}}
\newcommand{\mat}[1]{ {\bf #1} } 
\newcommand{\msun}{{\rm M}_{\odot}}

\newcommand{\grape}{{\small GRAPE}}
\newcommand{\gadget}{{\small GADGET}}
\newcommand{\bmath}{\bf}

\usepackage[small,bf]{caption}

\renewcommand{\thebibliography}[1]{
\section*{References}
 \list{}{\setlength{\leftmargin}{1.5em}
         \setlength{\itemsep}{-0.3em}
         \itemindent-\leftmargin}
}

\begin{document}

\date{}
\title{\vspace*{-2cm}GADGET: A code for collisionless and gasdynamical cosmological simulations}
\author{Volker Springel$^{1,2}$, Naoki Yoshida$^1$ and Simon D. M. White$^1$\\
{\normalsize $^1$Max-Planck-Institut f\"{u}r Astrophysik, Karl-Schwarzschild-Stra\ss{}e 1,
85740 Garching bei M\"{u}nchen, Germany}\vspace*{-0.5ex}\\
{\normalsize $^2$Harvard-Smithsonian Center for Astrophysics, 60 Garden Street, Cambridge, MA~02138, USA}\\
\vspace*{0.5cm}\ \\
\parbox{14cm} {\normalsize{\sffamily\bfseries Abstract}\newline We
describe the newly written code \gadget\ which is suitable both for
cosmological simulations of structure formation and for the simulation
of interacting galaxies. \gadget\ evolves self-gravitating
collisionless fluids with the traditional N-body approach, and a
collisional gas by smoothed particle hydrodynamics.  Along with the
serial version of the code, we discuss a parallel version that has
been designed to run on massively parallel supercomputers with
distributed memory. While both versions use a tree algorithm to
compute gravitational forces, the serial version of \gadget\ can
optionally employ the special-purpose hardware \grape\ instead of the
tree. Periodic boundary conditions are supported by means of an Ewald
summation technique.  The code uses individual and adaptive timesteps
for all particles, and it combines this with a scheme for dynamic tree
updates. Due to its Lagrangian nature, \gadget\ thus allows a very
large dynamic range to be bridged, both in space and time.  So far,
\gadget\ has been successfully used to run simulations with up to
7.5$\times 10^7$ particles, including cosmological studies of
large-scale structure formation, high-resolution simulations of the
formation of clusters of galaxies, as well as workstation-sized
problems of interacting galaxies.  In this study, we detail the
numerical algorithms employed, and show various tests of the code. We
publically release both the serial and the massively parallel version
of the code.
\vspace{1ex}\ 
}
}

\maketitle

\section{Introduction}

Numerical simulations of three-dimensional self-gravitating fluids
have become an indispensable tool in cosmology. They are now routinely
used to study the non-linear gravitational clustering of dark matter,
the formation of clusters of galaxies, the interactions of isolated
galaxies, and the evolution of the intergalactic gas. Without
numerical techniques the immense progress made in these fields would
have been nearly impossible, since analytic calculations are often
restricted to idealized problems of high symmetry, or to approximate
treatments of inherently nonlinear problems.

The advances in numerical simulations have become possible both by the
rapid growth of computer performance and by the implementation of ever
more sophisticated numerical algorithms.  The development of powerful
simulation codes still remains a primary task if one wants to take
full advantage of new computer technologies.

Early simulations \citep[][among
others]{Ho41,Pee70,Pre74,Wh76,Wh78,Aar79} largely employed the direct
summation method for the gravitational N-body problem, which remains
useful in collisional stellar dynamical systems, but it is inefficient
for large $N$ due to the rapid increase of its computational cost with
$N$.  A large number of groups have therefore developed N-body codes
for collisionless dynamics that compute the large-scale gravitational
field by means of Fourier techniques. These are the PM, P$^3$M, and
AP$^3$M codes \citep{Eas74,Hohl78,Hock81,Ef85,Cou91,Ber91,Ma98}.
Modern versions of these codes supplement the force computation on
scales below the mesh size with a direct summation, and/or they place
mesh refinements on highly clustered regions. Poisson's equation can
also be solved on a hierarchically refined mesh by means of
finite-difference relaxation methods, an approach taken in the ART
code by \citet{Kra97}.

An alternative to these schemes are the so-called tree algorithms,
pioneered by \citet{App81,App85}. Tree algorithms arrange particles in
a hierarchy of groups, and compute the gravitational field at a given
point by summing over multipole expansions of these groups. In this
way the computational cost of a complete force evaluation can be
reduced to a ${\cal O}(N\log N)$ scaling. The grouping itself can be
achieved in various ways, for example with Eulerian subdivisions of
space \citep{Ba86}, or with nearest-neighbour pairings
\citep{Press86,Je89}. A technique related to ordinary tree algorithms
is the fast multipole-method \citep[e.g.][]{Green87}, where multipole
expansions are carried out for the gravitational field in a region of
space.

While mesh-based codes are generally much faster for
close-to-homogeneous particle distributions, tree codes can adapt
flexibly to any clustering state without significant losses in speed.
This Lagrangian nature is a great advantage if a large dynamic range
in density needs to be covered. Here tree codes can outperform mesh
based algorithms.  In addition, tree codes are basically free from any
geometrical restrictions, and they can be easily combined with
integration schemes that advance particles on individual timesteps.

Recently, PM and tree solvers have been combined into hybrid Tree-PM
codes \citep{Xu95,Bag99,Bod99}. In this approach, the speed and
accuracy of the PM method for the long-range part of the gravitational
force are combined with a tree-computation of the short-range force.
This may be seen as a replacement of the direct summation PP part in
P$^3$M codes with a tree algorithm.  The Tree-PM technique is clearly
a promising new method, especially if large cosmological volumes with
strong clustering on small scales are studied.

Yet another approach to the N-body problem is provided by
special-purpose hardware like the \grape\ board
\citep{Mak90,Ito91,Fuk91,Mak93,Ebi93,Okam93,Fuk96,Ma97,Kaw2000}. It
consists of custom chips that compute gravitational forces by the
direct summation technique. By means of their enormous computational
speed they can considerably extend the range where direct summation
remains competitive with pure software solutions. A recent overview of
the family of \grape-boards is given by \citet{Hut99}.  The newest
generation of \grape\ technology, the \grape-6, will achieve a peak
performance of up to 100 TFlops \citep{Mak2000}, allowing direct
simulations of dense stellar systems with particle numbers approaching
$10^6$. Using sophisticated algorithms, \grape\ may also be combined
with P$^3$M \citep{Brieu95} or tree algorithms
\citep{Fuk91,Mak91,At97} to maintain its high computational speed even
for much larger particle numbers.

In recent years, collisionless dynamics has also been coupled to gas
dynamics, allowing a more direct link to observable quantities.
Traditionally, hydrodynamical simulations have usually employed some
kind of mesh to represent the dynamical quantities of the fluid. While
a particular strength of these codes is their ability to accurately
resolve shocks, the mesh also imposes restrictions on the geometry of
the problem, and onto the dynamic range of spatial scales that can be
simulated.  New adaptive mesh refinement codes \citep{No98,Kl98} have
been developed to provide a solution to this problem.

In cosmological applications, it is often sufficient to describe the
gas by smoothed particle hydrodynamics (SPH), as invented by
\citet{Lu77} and \citet{Gi77}. The particle-based SPH is extremely
flexible in its ability to adapt to any given geometry. Moreover, its
Lagrangian nature allows a locally changing resolution that
`automatically' follows the local mass density. This convenient
feature helps to save computing time by focusing the computational
effort on those regions that have the largest gas concentrations.
Furthermore, SPH ties naturally into the N-body approach for
self-gravity, and can be easily implemented in three dimensions.

These advantages have led a number of authors to develop SPH codes for
applications in cosmology. Among them are {\small TREESPH}
\citep{He89,Ka96}, {\small GRAPESPH} \citep{St96}, {\small HYDRA}
\citep{Cou95,Pe97}, and codes by \citet{Ev88,Na93,Hu97,Da97,Ca98}.
See \cite{Ka94} and \cite{Fr98} for a comparison of many of these
cosmological hydrodynamic codes.

In this paper we describe our simulation code \gadget\ ({\bf GA}laxies
with {\bf D}ark matter and {\bf G}as int{\bf E}rac{\bf T}), which can
be used both for studies of isolated self-gravitating systems
including gas, or for cosmological N-body/SPH simulations. We have
developed two versions of this code, a serial workstation version, and
a version for massively parallel supercomputers with distributed
memory.  The workstation code uses either a tree algorithm for the
self-gravity, or the special-purpose hardware \grape, if available.
The parallel version works with a tree only. Note that in principle
several \grape\ boards, each connected to a separate host computer,
can be combined to work as a large parallel machine, but this
possibility is not implemented in the parallel code yet. While the
serial code largely follows known algorithmic techniques, we employ a
novel parallelization strategy in the parallel version.

A particular emphasis of our work has been on the use of a time
integration scheme with individual and adaptive particle timesteps,
and on the elimination of sources of overhead both in the serial and
parallel code under conditions of large dynamic range in
timestep. Such conditions occur in dissipative gas-dynamical
simulations of galaxy formation, but also in high-resolution
simulations of cold dark matter.  The code allows the usage of
different timestep criteria and cell-opening criteria, and it can be
comfortably applied to a wide range of applications, including
cosmological simulations (with or without periodic boundaries),
simulations of isolated or interacting galaxies, and studies of the
intergalactic medium.

We thus think that \gadget\ is a very flexible code that avoids
obvious intrinsic restrictions for the dynamic range of the problems
that can be addressed with it.  In this methods-paper, we describe the
algorithmic choices made in \gadget\, which we release in its parallel
and serial versions on the internet\footnote{{\scriptsize GADGET}'s
web-site is:\\ {\tt http://www.mpa-garching.mpg.de/gadget}}, hoping
that it will be useful for people working on cosmological simulations,
and that it will stimulate code development efforts and further
code-sharing in the community.

This paper is structured as follows.  In Section \ref{secphys}, we
give a brief summary of the implemented physics. In Section
\ref{secgrav}, we discuss the computation of the gravitational force
both with a tree algorithm, and with \grape. We then describe our
specific implementation of SPH in Section \ref{secsph}, and we discuss
our time integration scheme in Section \ref{sectime}.  The
parallelization of the code is described in Section \ref{secpara}, and
tests of the code are presented in Section \ref{secres}. Finally, we
summarize in Section~\ref{secdis}.

\section{Implemented physics}\label{secphys}

\subsection{Collisionless dynamics and gravity}

Dark matter and stars are modeled as self-gravitating collisionless
fluids, i.e.\ they fulfill the collisionless Boltzmann equation (CBE)
\be
\frac{\dd f}{\dd t}
\equiv\frac{\partial f}{\partial t}+ \vec{v}\frac{\partial f}{\partial \vec{x}}
-\frac{\partial\Phi}{\partial \vec{r}}\frac{\partial f}{\partial\vec{v}}=0,
\ee
where the self-consistent potential $\Phi$ is the solution of
Poisson's equation
\be
\nabla^2\Phi(\vec{r},t)=4\pi G \int f(\vec{r},\vec{v},t) \dd \vec{v},
\ee
and $f(\vec{r},\vec{v},t)$ is the mass density in single-particle
phase-space.  It is very difficult to solve this coupled system of
equations directly with finite difference methods. Instead, we will
follow the common {N-body} approach, where the phase fluid is
represented by $N$ particles which are integrated along the
characteristic curves of the CBE. In essence, this is a Monte Carlo
approach whose accuracy depends crucially on a sufficiently high
number of particles.

The N-body problem is thus the task of following Newton's equations of
motion for a large number of particles under their own
self-gravity. Note that we will introduce a softening into the
gravitational potential at small separations.  This is necessary to
suppress large-angle scattering in two-body collisions and effectively
introduces a lower spatial resolution cut-off. For a given softening
length, it is important to choose the particle number large enough
such that relaxation effects due to two-body encounters are suppressed
sufficiently, otherwise the N-body system provides no faithful model
for a collisionless system.  Note that the optimum choice of softening
length as a function of particle density is an issue that is still
actively discussed in the literature \citep[e.g.][]{Spl98,Rom98,At99}.

\subsection{Gasdynamics}

A simple description of the intergalactic medium (IGM), or the
interstellar medium (ISM), may be obtained by modeling it as an ideal,
inviscid gas. The gas is then governed by the continuity equation
\be
\frac{\dd \rho}{\dd t} + \rho \nabla\cdot\vec{v} =0 ,
\ee
and the Euler equation
\be
\frac{\dd v}{\dd t}= -\frac{\nabla P}{\rho} -\nabla \Phi .
\ee
Further, the thermal energy $u$ per unit mass evolves according to the
first law of thermodynamics, viz.\
\be
\frac{\dd u}{\dd t} =
-\frac{P}{\rho}\nabla\cdot \vec{v}
-\frac{\Lambda(u,\rho)}{\rho}.
\ee
Here we used Lagrangian time derivatives, i.e.
\be
\frac{\dd} { \dd t} =\frac{\partial}{\partial t} + \vec{v}\cdot\nabla ,
\ee
and we allowed for a piece of `extra' physics in form of the {\it
cooling function} $\Lambda(u,\rho)$, describing external sinks or
sources of heat for the gas.

For a simple ideal gas, the equation of state is
\be
P=(\gamma-1)\rho u,
\ee
where $\gamma$ is the adiabatic exponent.  We usually take
$\gamma=5/3$, appropriate for a mono-atomic ideal gas.  The adiabatic
sound speed $c$ of this gas is $c^2= {\gamma P}/{\rho}$.
\vspace*{0.5cm}\ \\

\section{Gravitational forces}  \label{secgrav} 	

\subsection{Tree algorithm} \label{sectree}

\begin{figure*}
\bc
\rotatebox{90}{\resizebox{4cm}{!}{\includegraphics{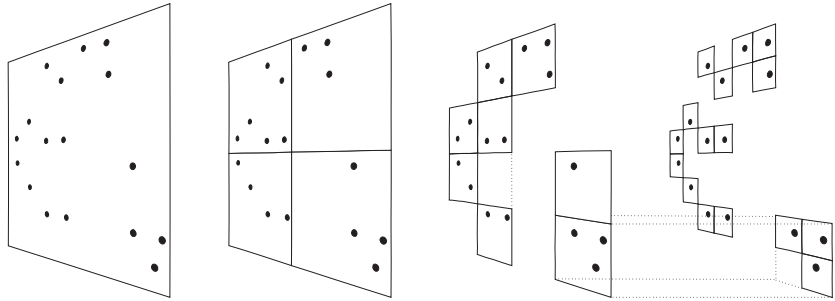}}}\\
\caption{Schematic illustration of the Barnes \& Hut oct-tree in two
dimensions.  The particles are first enclosed in a square (root
node). This square is then iteratively subdivided in four squares of
half the size, until exactly one particle is left in each final square
(leaves of the tree). In the resulting tree structure, each square can
be progenitor of up to four siblings. Note that empty squares need not
to be stored.
\label{figBH}}
\ec
\end{figure*}

An alternative to Fourier techniques, or to direct summation, are the
so-called tree methods. In these schemes, the particles are arranged
in a hierarchy of groups. When the force on a particular particle is
computed, the force exerted by distant groups is approximated by their
lowest multipole moments.  In this way, the computational cost for a
complete force evaluation can be reduced to order ${\cal O}(N\log N)$
\citep{App85}.  The forces become more accurate if the multipole
expansion is carried out to higher order, but eventually the
increasing cost of evaluating higher moments makes it more efficient
to terminate the multipole expansion and rather use a larger number of
smaller tree nodes to achieve a desired force accuracy
\citep{McMill93}.  We will follow the common compromise to terminate
the expansion after quadrupole moments have been included.

We employ the \citet[henceforth BH]{Ba86} tree construction in this
work.  In this scheme, the computational domain is hierarchically
partitioned into a sequence of cubes, where each cube contains eight
siblings, each with half the side-length of the parent cube. These
cubes form the nodes of an oct-tree structure.  The tree is
constructed such that each node (cube) contains either exactly one
particle, or is progenitor to further nodes, in which case the node
carries the monopole and quadrupole moments of all the particles that
lie inside its cube.  A schematic illustration of the BH tree is shown
in Figure \ref{figBH}.

A force computation then proceeds by walking the tree, and summing up
appropriate force contributions from tree nodes. In the standard BH
tree walk, the multipole expansion of a node of size $l$ is used only
if 
\be r>\frac{l}{\theta} \, ,
\label{eqopen}
\ee where $r$ is the distance of the point of reference to the
center-of-mass of the cell and $\theta$ is a prescribed accuracy
parameter.  If a node fulfills the criterion (\ref{eqopen}), the tree
walk along this branch can be terminated, otherwise it is `opened',
and the walk is continued with all its siblings. For smaller values of
the opening angle, the forces will in general become more accurate,
but also more costly to compute. One can try to modify the opening
criterion (\ref{eqopen}) to obtain higher efficiency, i.e.~higher
accuracy at a given length of the interaction list, something that we
will discuss in more detail in Section~\ref{newopsec}.

A technical difficulty arises when the gravity is softened.  In
regions of high particle density (e.g.~centers of dark haloes, or cold
dense gas knots in dissipative simulations), it can happen that nodes
fulfill equation (\ref{eqopen}), and simultaneously one has $r<h$,
where $h$ is the gravitational softening length. In this situation,
one formally needs the multipole moments of the {\em softened}
gravitational field. One can work around this situation by opening
nodes always for $r<h$, but this can slow down the code significantly
if regions of very high particle density occur. Another solution is to
use the proper multipole expansion for the softened potential, which
we here discuss for definiteness.  We want to approximate the
potential at $\vec{r}$ due to a (distant) bunch of particles with
masses $m_i$ and coordinates $\vec{x}_i$. We use a spline-softened
force law, hence the exact potential of the particle group is \be
\Phi(\vec{r}) =-G \sum_k m_k\, g \left( \left| \vec{x}_k - \vec{r}
\right| \right), \ee where the function $g(r)$ describes the softened
force law. For Newtonian gravity we have $g(r)=1/r$, while the spline
softened gravity with softening length $h$ gives rise to \be
g(r)=-\frac{1}{h}W_2\left(\frac{r}{h}\right).  \ee The function
$W_2(u)$ is given in the Appendix. It arises by replacing the force
due to a point mass $m$ with the force exerted by the mass
distribution $\rho(\vec{r})=m W(\vec{r};h)$, where we take $W(r;h)$ to
be the normalized spline kernel used in the SPH formalism.  The spline
softening has the advantage that the force becomes exactly Newtonian
for $r>h$, while some other possible force laws, like the Plummer
softening, converge relatively slowly to Newton's law.

Let $\vec{s}$ be the center-of-mass, and $M$ the total mass 
of the particles. Further we define $\vec{y}\equiv \vec{r}-\vec{s}$.
The potential may then be expanded in a multipole series assuming
$|\vec{y}| \gg |\vec{x}_k-\vec{s}|$.  Up to quadrupole order, this
results in 
\bea
\label{eqq1}
\Phi(\vec{r}) 
& 
= 
& 
-G  \bigg\{ 
M g(y)   
\\ & & 
\left.
\;\;\;\;\;\;\;\;\; 
+\frac{1}{2}
\vec{y}^T 
\left[ \frac{g''(y)}{y^2} \mat{Q} + \frac{g'(y)}{y^3}(\mat{P}-\mat{Q}) \right]
\vec{y}
\right\} .
\nonumber
\eea
Here we have introduced the tensors
\be
\mat{Q} =\sum_k m_k ( \vec{x}_k - \vec{s}) (\vec{x}_k - \vec{s})^T 
= \sum_k m_k  \vec{x}_k \, \vec{x}_k^T  -  M \vec{s}\,\vec{s}^T ,
\ee
and
\be
\mat{P} = \mat{I} \sum_k m_k ( \vec{x}_k - \vec{s})^2 
= \mat{I}\left[ \sum_k m_k  \vec{x}_k^2   -  M \vec{s}^2\right],
\ee
where $\mat{I}$ is the unit matrix.
Note that for Newtonian gravity, equation (\ref{eqq1}) reduces to the
more familiar form
\be
\Phi(\vec{r})
=-G  \left[
\frac{M}{y}
+\frac{1}{2}
\vec{y}^T 
\frac{\,\, 3\mat{Q} -\mat{P}\,\, }{y^5}
\vec{y}
\right].
\label{eqq2}
\ee Finally, the quadrupole approximation of the softened
gravitational field is given by 
\bea
\label{eqqF1}
\vec{f}(\vec{r})
=-\nabla \Phi=G\Big\{
M g_1(y) \vec{y}
+ g_2(y)  \mat{Q}\vec{y} 
\nonumber   \\
\hspace{1.4cm}\left. 
+ \frac{1}{2} g_3(y)  \left( \vec{y}^T  \mat{Q} \vec{y} \right) \vec{y}
+ \frac{1}{2} g_4(y) \mat{P} \vec{y}
\right\} .
\eea
Here we introduced the functions $g_1(y)$, $g_2(y)$, $g_3(y)$, and
$g_4(y)$ as convenient abbreviations. Their definition is given in the
Appendix. In the Newtonian case, this simplifies to
\be
\label{eqqF2}
\hspace{-0.2cm}\vec{f}(\vec{r})
=G\left\{
- \frac{M}{y^3} \vec{y}
+  \frac{3 \mat{Q}}{y^5} \vec{y}
- \frac{15}{2} \frac{\vec{y}^T\vec{Q}\vec{y}}{y^7} \vec{y}
+\frac{3}{2}\frac{\mat{P}}{y^5} \vec{y}
\right\} .
\ee

Note that although equation (\ref{eqqF1}) looks rather cumbersome, its
actual numerical computation is only marginally more costly than that
of the Newtonian form (\ref{eqqF2}) because all factors involving
$g(y)$ and derivatives thereof can be tabulated for later use in
repeated force calculations.

\subsection{Tree construction and tree walks}

The tree construction can be done by inserting the particles one after
the other in the tree. Once the grouping is completed, the multipole
moments of each node can be recursively computed from the moments of
its daughter nodes \citep{McMill93}.

In order to reduce the storage requirements for tree nodes, we use
single-precision floating point numbers to store node properties.  The
precision of the resulting forces is still fully sufficient for
collisionless dynamics as long as the node properties are calculated
accurately enough. In the recursive calculation, node properties will
be computed from nodes that are already stored in single precision.
When the particle number becomes very large (note that more than 10
million particles can be used in single objects like clusters these
days), loss of sufficient precision can then result for certain
particle distributions. In order to avoid this problem, \gadget\
optionally uses an alternative method to compute the node properties.
In this method, a link-list structure is used to access all of the
particles represented by each tree node, allowing a computation of the
node properties in double-precision and a storage of the results in
single-precision. While this technique guarantees that node properties
are accurate up to a relative error of about $10^{-6}$, it is also
slower than the recursive computation, because it requires of order
$\cal{O}(N\log N)$ operations, while the recursive method is only of
order $\cal{O}(N)$.

The tree-construction can be considered very fast in both cases,
because the time spent for it is negligible compared to a complete
force walk for all particles.  However, in the individual time
integration scheme only a small fraction of all particles may require
a force walk at each given timestep. If this fraction drops below
$\sim 1$ per cent, a full reconstruction of the tree can take as much
time as the force walk itself. Fortunately, most of this tree
construction time can be eliminated by dynamic tree updates
\citep{McMill93}, which we discuss in more detail in Section
\ref{sectime}.  The most time consuming routine in the code will then
always remain the tree walk, and optimizing it can considerably speed
up tree codes. Interestingly, in the grouping technique of
\citet{Ba90}, the speed of the gravitational force computation can be
increased by performing a common tree-walk for a localized group of
particles. Even though the average length of the interaction list for
each particles becomes larger in this way, this can be offset by
saving some of the tree-walk overhead, and by improved cache
utilization. Unfortunately, this advantage is not easily kept if
individual timesteps are used, where only a small fraction of the
particles are active, so we do not use grouping.

\gadget\ allows different gravitational softenings for particles of
different `type'. In order to guarantee momentum conservation, this
requires a symmetrization of the force when particles with different
softening lengths interact.  We symmetrize the softenings
in the form \be h=\max(h_i,h_j).  \ee However, the usage of different
softening lengths leads to complications for softened tree nodes,
because strictly speaking, the multipole expansion is only valid if
all the particles in the node have the same softening.  \gadget\
solves this problem by constructing separate trees for each species of
particles with different softening.  As long as these species are more
or less spatially separated (e.g.\ dark halo, stellar disk, and
stellar bulge in simulations of interacting galaxies), no severe
performance penalty results. However, this is different if the fluids
are spatially well `mixed'.  Here a single tree would result in higher
performance of the gravity computation, so it is advisable to choose a
single softening in this case. Note that for SPH particles we
nevertheless always create a separate tree to allow its use for a fast
neighbour search, as will be discussed below.

\subsection{Cell-opening criterion\label{newopsec}}

The accuracy of the force resulting from a tree walk depends
sensitively on the criterion used to decide whether the multipole
approximation for a given node is acceptable, or whether the node has
to be `opened' for further refinement.  The standard BH opening
criterion tries to limit the relative error of every particle-node
interaction by comparing a rough estimate of the size of the
quadrupole term, $\sim Ml^2/r^4$, with the size of the monopole term,
$\sim M/r^2$. The result is the purely geometrical criterion of
equation (\ref{eqopen}).

However, as \citet{Sal94} have pointed out, the worst-case behaviour
of the BH criterion for commonly employed opening angles is somewhat
worrying.  Although typically very rare in real astrophysical
simulations, the geometrical criterion (\ref{eqopen}) can then
sometimes lead to very large force errors. In order to cure this
problem, a number of modifications of the cell-opening criterion have
been proposed. For example, \citet{Du96} have used the simple
modification $ r>{l}/{\theta} + \delta$, where the quantity $\delta$
gives the distance of the geometric center of the cell to its
center-of-mass. This provides protection against pathological cases
where the center-of-mass lies close to an edge of a cell.

Such modifications can help to reduce the rate at which large force
errors occur, but they usually do not help to deal with another
problem that arises for geometric opening criteria in the context of
cosmological simulations at high redshift. Here, the density field is
very close to being homogeneous and the peculiar accelerations are
small. For a tree algorithm this is a surprisingly tough problem,
because the tree code always has to sum up partial forces from {\em
all} the mass in a simulation. Small net forces at high $z$ then arise
in a delicate cancellation process between relatively large partial
forces. If a partial force is indeed much larger than the net force,
even a small relative error in it is enough to result in a large
relative error of the net force. For an unclustered particle
distribution, the BH criterion therefore requires a much smaller value
of the opening angle than for a clustered one in order to achieve a
similar level of force accuracy.  Also note that in a cosmological
simulation the absolute sizes of forces between a given particle and
tree-nodes of a certain opening angle can vary by many orders of
magnitude. In this situation, the purely geometrical BH criterion may
end up investing a lot of computational effort for the evaluation of
all partial forces to the same relative accuracy, irrespective of the
actual size of each partial force and the size of the absolute error
thus induced. It would be better to invest more computational effort
in regions that provide most of the force on the particle and less in
regions whose mass content is unimportant for the total force.

As suggested by \citet{Sal94}, one may therefore try to devise a
cell-opening criterion that limits the absolute error in every
cell-particle interaction. In principle, one can use analytic error
bounds \citep{Sal94} to obtain a suitable cell-opening criterion, but
the evaluation of the relevant expressions can consume significant
amounts of CPU time.

Our approach to a new opening criterion is less stringent. Assume the
absolute size of the true total force is already known before the tree
walk. In the present code, we will use the acceleration of the
previous timestep as a handy approximate value for that. We will now
require that the estimated error of an acceptable multipole
approximation is some small fraction of this total force.  Since we
truncate the multipole expansion at quadrupole order, the octupole
moment will in general be the largest term in the neglected part of
the series, except when the mass distribution in the cubical cell is
close to being homogeneous. For a homogeneous cube the octupole moment
vanishes by symmetry \citep{BaH89}, such that the hexadecapole moment
forms the leading term.  We may very roughly estimate the size of
these terms as $\sim M/r^2 (l/r)^3$, or $\sim M/r^2 (l/r)^4$,
respectively, and take this as a rough estimate of the size of the
truncation error. We can then require that this error should not
exceed some fraction $\alpha$ of the total force on the particle,
where the latter is estimated from the previous timestep. Assuming the
octupole scaling, a tree-node has then to be opened if $M\, l^3>
\alpha |\vec{a}_{\rm old}| r^5$. However, we have found that in
practice the opening criterion \be M\, l^4> \alpha |\vec{a}_{\rm old}|
r^6 \, \label{eqnewcrit} \ee provides still better performance in the
sense that it produces forces that are more accurate at a given
computational expense. It is also somewhat cheaper to evaluate during
the tree-walk, because $r^6$ is simpler to compute than $r^5$, which
requires the evaluation of a root of the squared node distance.
The criterion (\ref{eqnewcrit}) does not suffer from the high-$z$
problem discussed above, because the same value of $\alpha$ produces a
comparable force accuracy, independent of the clustering state of the
material. However, we still need to compute the very first force using
the BH criterion. In Section~\ref{newopsecresults}, we will show some
quantitative measurements of the relative performance of the two
criteria, and compare it to the optimum cell-opening strategy.

Note that the criterion (\ref{eqnewcrit}) is not completely safe from
worst-case force errors either. In particular, such errors can occur
for opening angles so large that the point of force evaluation falls
into the node itself.  If this happens, no upper bound on the force
error can be guaranteed \citep{Sal94}. As an option to the code, we
therefore combine the opening criterion (\ref{eqnewcrit}) with the
requirement that the point of reference may not lie inside the node
itself. We formulate this additional constraint in terms of $r>b_{\rm
max}$, where $b_{\rm max}$ is the maximum distance of the
center-of-mass from any point in the cell. This additional geometrical
constraint provides a very conservative control of force errors if
this is needed, but increases the number of opened cells.

\subsection{Special purpose hardware}	

An alternative to software solutions to the $N^2$-bottleneck of
self-gravity is provided by the \grape\ (GRAvity PipE) special-purpose
hardware.  It is designed to solve the gravitational N-body problem in
a direct summation approach by means of its superior computational
speed.  The latter is achieved with custom chips that compute the
gravitational force with a hardwired Plummer force law.  The
Plummer-potential of \grape\ takes the form \be
\Phi(\vec{r})=-G\sum_j\frac{m_j}{\left(|\vec{r}-\vec{r}_j|^2 +
\epsilon^2\right)^{\frac{1}{2}}}.  \ee

As an example, the \grape-3A boards installed at the MPA in 1998 have
40 N-body integrator chips in total with an approximate peak
performance of 25 GFlops. Recently, newer generations of \grape\
boards have achieved even higher computational speeds.  In fact, with
the \grape-4 the 1 TFlop barrier was broken \citep{Ma97}, and even
faster special-purpose machines are in preparation
\citep{Hut99,Mak2000}. The most recent generation, {\grape-6}, can not
only compute accelerations, but also its first and second time
derivatives. Together with the capability to perform particle
predictions, these machines are ideal for high-order Hermite
integration schemes applied in simulations of collisional systems like
star clusters.  However, our present code is only adapted to the
somewhat older \grape-3 \citep{Okam93}, and the following discussion
is limited to it.

The \grape-3A boards are connected to an ordinary workstation via a
VME or PCI interface. The boards consist of memory chips that can hold
up to 131072 particle coordinates, and of integrator chips that can
compute the forces exerted by these particles for 40 positions in
parallel. Higher particle numbers can be processed by splitting them
up in sufficiently small groups.  In addition to the gravitational
force, the \grape\ board returns the potential, and a list of
neighbours for the 40 positions within search radii $h_i$ specified by
the user. This latter feature makes \grape\ attractive also for SPH
calculations.

The parts of our code that use \grape\ have benefited from the code
{\small GRAPESPH} by \citet{St96}, and are similar to it.  In short,
the usage of \grape\ proceeds as follows.  For the force computation,
the particle coordinates are first loaded onto the \grape\ board, then
\gadget\ calls \grape\ repeatedly to compute the force for up to 40
positions in parallel.  The communication with \grape\ is done by
means of a convenient software interface in C. \grape\ can also
provide lists of nearest neighbours. For SPH-particles, \gadget\
computes the gravitational force and the interaction list in just one
call of \grape. The host computer then still does the rest of the
work, i.e.\ it advances the particles, and computes the hydrodynamical
forces.

In practice, there are some technical complications when one works
with \grape-3.  In order to achieve high computational speed, the
\grape-3 hardware works internally with special fixed-point formats
for positions, accelerations and masses.  This results in a reduced
dynamic range compared to standard IEEE floating point arithmetic.  In
particular, one needs to specify a minimum length scale $d_{\rm min}$
and a minimum mass scale $m_{\rm min}$ when working with \grape. The
spatial dynamic range is then given by $d_{\rm min}[-2^{18};2^{18}]$
and the mass range is $m_{\rm min}[1;64{\epsilon}/{d_{\rm min}}]$
\citep{St96}.

While the communication time with \grape\ scales proportional to the
particle number $N$, the actual force computation of \grape\ is still
an ${\cal O}(N^2)$-algorithm, because the \grape\ board implements a
direct summation approach to the gravitational N-body problem.  This
implies that for very large particle number a tree code running on the
workstation alone will eventually catch up and outperform the
combination of workstation and \grape.  For our current set-up at MPA
this break-even point is about at 300000 particles.

However, it is also possible to combine \grape\ with a tree algorithm
\citep{Fuk91,Mak91,At97,Kaw2000}, for example by exporting tree nodes
instead of particles in an appropriate way. Such a combination of
tree+\grape\ scales as ${\cal O}(N\log N)$ and is able to outperform
pure software solutions even for large $N$.

\section{Smoothed particle hydrodynamics} \label{secsph}

SPH is a powerful Lagrangian technique to solve hydrodynamical
problems with an ease that is unmatched by grid based fluid solvers
\citep[see][for an excellent review]{Mo92}. In particular, SPH is very
well suited for three-dimensional astrophysical problems that do not
crucially rely on accurately resolved shock fronts.

Unlike other numerical approaches for hydrodynamics, the SPH equations
do not take a unique form. Instead, many formally different versions
of them can be derived. Furthermore, a large variety of recipes for
specific implementations of force symmetrization, determinations of
smoothing lengths, and artificial viscosity, have been described. Some
of these choices are crucial for the accuracy and efficiency of the
SPH implementation, others are only of minor importance. See the
recent work by \citet{Th98} and \citet{Lo98} for a discussion of the
relative performance of some of these possibilities.  Below we give a
summary of the specific SPH implementation we use.

\subsection{Basic equations}

The computation of the hydrodynamic force and the rate of change of
internal energy proceeds in two phases.  In the first phase, new
smoothing lengths $h_i$ are determined for the {\em active} particles
(these are the ones that need a force update at the current timestep,
see below), and for each of them, the neighbouring particles inside
their respective smoothing radii are found. The Lagrangian nature of
SPH arises when this number of neighbours is kept either exactly, or
at least roughly, constant.  This is achieved by varying the smoothing
length $h_i$ of each particle accordingly. The $h_i$ thus adjust to
the local particle density adaptively, leading to a constant mass
resolution independent of the density of the flow.  \citet{Ne94} argue
that it is actually best to keep the number of neighbours exactly
constant, resulting in the lowest level of noise in SPH estimates of
fluid quantities, and in the best conservation of energy. In practice,
similarly good results are obtained if the fluctuations in neighbour
number remain very small.  In the serial version of \gadget\ we keep
the number of neighbours fixed, whereas it is allowed to vary in a
small band in the parallel code.

Having found the neighbours, we compute the density of the active
particles as \be \rho_i=\sum_{j=1}^N m_j W(\vec{r}_{ij};h_i), \ee
where $\vec{r}_{ij}\equiv \vec{r}_i - \vec{r}_j$, and we compute a new
estimate of divergence and vorticity as 
\be \rho_i
(\nabla\cdot\vec{v})_i =\sum_j m_j (\vec{v}_j-\vec{v}_i)\nabla_i
W(\vec{r}_{ij};h_i) , 
\ee 
\be \rho_i (\nabla\times\vec{v})_i =\sum_j
m_j (\vec{v}_i-\vec{v}_j)\times \nabla_i W(\vec{r}_{ij};h_i) .  
\ee
Here we employ the {\em gather} formulation for adaptive smoothing
\citep{He89}.

For the {\em passive} particles, values for density, internal energy,
and smoothing length are predicted at the current time based on the
values of the last update of those particles (see Section
\ref{sectime}).  Finally, the pressure of the particles is set to
$P_i=(\gamma-1)\rho_i u_i$.

In the second phase, the actual forces are computed.  Here we
symmetrize the kernels of {\em gather} and {\em scatter} formulations
as in \citet{He89}.  We compute the gasdynamical accelerations as 
\bea
\label{eqsphA}
\vec{a}_i^{\rm gas}  =  -\left(\frac{\nabla P}{\rho}\right)_i + \vec{a}_i^{\rm visc}
 = 
-\sum_j m_j\left( \frac{P_i}{\rho_i^2}+\frac{P_j}{\rho_j^2}+
\tilde\Pi_{ij}\right)  \nonumber\hspace*{-1cm}\  & &\\
\left[ \frac{1}{2} \nabla_i W(\vec{r}_{ij};h_{i}) + \frac{1}{2}
\nabla_i W(\vec{r}_{ij};h_{j})
\right], & &
\eea
and the change of the internal energy as
\bea
\label{eqsphB}
\frac{\dd u_i}{\dd t}=
\frac{1}{2}\sum_j m_j\left( 
\frac{P_i}{\rho_i^2}+\frac{P_j}{\rho_j^2}+
\tilde\Pi_{ij}\right)(\vec{v}_i - \vec{v}_j)\, \hspace*{0cm}& & \nonumber \\
\left[ \frac{1}{2} \nabla_i W(\vec{r}_{ij};h_{i}) + \frac{1}{2}
\nabla_i W(\vec{r}_{ij};h_{j})
\right]. & &
\eea
Instead of symmetrizing the pressure terms with an arithmetic mean,
the code can also be used
with a geometric mean according to
\be
\frac{P_i}{\rho_i^2}+\frac{P_j}{\rho_j^2} \;\; \longrightarrow\;\;
2\frac{\sqrt{P_i P_j}}{\rho_i\rho_j}.
\ee
This may be slightly more robust in certain situations \citep{He89}.
The artificial viscosity $\tilde\Pi_{ij}$ is taken to be
\be
\label{eqvisc0}
\tilde\Pi_{ij}=\frac{1}{2}(f_i+f_j)\Pi_{ij},
\ee
with
\be
\label{eqvisc}
\Pi_{ij}=\left\{
\begin{tabular}{cl}
${\left[-\alpha c_{ij} \mu_{ij} +2\alpha \mu_{ij}^2\right]}/{\rho_{ij}}$ & \mbox{if
$\vec{v}_{ij}\cdot\vec{r}_{ij}<0$} \\
0 & \mbox{otherwise}, \\
\end{tabular}
\right.
\ee
where
\be
f_i=\frac{|(\nabla\cdot\vec{v})_i|}{|(\nabla\cdot\vec{v})_i|+|(\nabla\times\vec{v})_i|},
\ee
and
\be
\mu_{ij}=\frac{h_{ij}(\vec{v}_i-\vec{v}_j)(\vec{r}_i-\vec{r}_j) }
{\left|\vec{r}_i-\vec{r}_j\right|^2 + \epsilon h_{ij}^2} .
\ee
This form of artificial viscosity is the shear-reduced version
\citep{Bal95,St96} of the `standard' \citet{Mo83} artificial
viscosity. Recent studies \citep{Lo98,Th98} that test SPH
implementations endorse it.

In equations (\ref{eqsphA}) and (\ref{eqsphB}), a given SPH particle
$i$ will interact with a particle $j$ whenever $|\vec{r}_{ij}|<h_i$
or $|\vec{r}_{ij}|<h_j$. Standard search techniques can relatively easily
find all neighbours of particle $i$ inside a sphere of radius $h_i$,
but making sure that one really finds all interacting pairs in the
case $h_j>h_i$ is slightly more tricky. One solution to this problem is to
simply find all neighbours of $i$ inside $h_i$, and to consider the
force components
\be \vec{f}_{ij} =
-m_i m_j\left( \frac{P_i}{\rho_i^2}+\frac{P_j}{\rho_j^2}+
\tilde\Pi_{ij}\right) \frac{1}{2} \nabla_i W(\vec{r}_{ij};h_{i}) .
\ee  
If we add $\vec{f}_{ij}$ to the force on $i$, and $-\vec{f}_{ij}$ to
the force on $j$, the sum of equation (\ref{eqsphA}) is reproduced,
and the momentum conservation is manifest. This also holds for the
internal energy. Unfortunately, this only works if all particles are
active. In an individual timestep scheme, we therefore need an
efficient way to find all the neighbours of particle $i$ in the above
sense, and we discuss our algorithm for doing this below.

\subsection{Neighbour search}

In SPH, a basic task is to find the nearest neighbours of each SPH
particle to construct its interaction list. Specifically, in the
implementation we have chosen we need to find all particles closer
than a search radius $h_i$ in order to estimate the density, and one
needs all particles with $|\vec{r}_{ij}|< \max(h_i,h_j)$ for the
estimation of hydrodynamical forces.  Similar to gravity, the naive
solution that checks the distance of {\em all} particle pairs is an
${\cal O}(N^2)$ algorithm which slows down prohibitively for large
particle numbers. Fortunately, there are faster search algorithms.

When the particle distribution is approximately homogeneous, perhaps
the fastest algorithms work with a search grid that has a cell size
somewhat smaller than the search radius. The particles are then first
coarse-binned onto this search grid, and link-lists are established
that quickly deliver only those particles that lie in a specific cell
of the coarse grid. The neighbour search proceeds then by {\em range
searching}; only those mesh cells that have a spatial overlap with the
search range have to be opened.

For highly clustered particle distributions and varying search ranges
$h_i$, the above approach quickly degrades, since the mesh chosen for
the coarse grid has not the optimum size for all particles.  A more
flexible alternative is to employ a geometric search tree. For this
purpose, a tree with a structure just like the BH oct-tree can be
employed, \citet{He89} were the first to use the gravity tree for this
purpose.  In \gadget\ we use the same strategy and perform a neighbour
search by walking the tree. A cell is `opened' (i.e.\ further
followed) if it has a spatial overlap with the rectangular search
range. Testing for such an overlap is faster with a rectangular search
range than with a spherical one, so we inscribe the spherical search
region into a little cube for the purpose of this walk.  If one
arrives at a cell with only one particle, this is added to the
interaction list if it lies inside the search radius.  We also
terminate a tree walk along a branch, if the cell lies {\em
completely} inside the search range. Then all the particles in the
cell can be added to the interaction list, without checking any of
them for overlap with the search range any more. The particles in the
cell can be retrieved quickly by means of a link-list, which can be
constructed along with the tree and allows a retrieval of all the
particles that lie inside a given cell, just as it is possible in the
coarse-binning approach. Since this short-cut reduces the length of
the tree walk and the number of required checks for range overlap, the
speed of the algorithm is increased by a significant amount.

With a slight modification of the tree walk, one can also find all
particles with $|\vec{r}_{ij}|<\max(h_i,h_j)$. For this purpose, we
store in each tree node the maximum SPH smoothing length occurring
among its particles. The test for overlap is then simply done between
a cube of side-length $\max(h_i, h_{\rm node})$ centered on the
particle $i$ and the node itself, where $h_{\rm node}$ is the maximum
smoothing length among the particles of the node.

There remains the task to keep the number of neighbours around a given
SPH particle approximately (or exactly) constant.  We solve this by
predicting a value $\tilde{h}_i$ for the smoothing length based on the
length $h_i$ of the previous timestep, the actual number of neighbours
$N_i$ at that timestep, and the local velocity divergence: \be
\tilde{h}_i = \frac{1}{2} h_i^{\rm (old)}\left[1+\left(\frac{N_{\rm
s}}{N_i}\right)^{1/3}\right] + \dot{h}_i\,\Delta t , \ee where
$\dot{h}_i=\frac{1}{3}h_i (\nabla\cdot\vec{v})_i$, and $N_{\rm s}$ is
the desired number of neighbours.  A similar form for updating the
smoothing lengths has been used by \citet{He89}, see also \citet{Th98}
for a discussion of alternative choices.  The term in brackets tries
to bring the number of neighbours back to the desired value if $N_i$
deviates from it.  Should the resulting number of neighbours
nevertheless fall outside a prescribed range of tolerance, we
iteratively adjust ${h}_i$ until the number of neighbours is again
brought back to the desired range.  Optionally, our code allows the
user to impose a minimum smoothing length for SPH, typically chosen as
some fraction of the gravitational softening length. A larger number
of neighbours than $N_{\rm s}$ is allowed to occur if $h_i$ takes on
this minimum value.

One may also decide to keep the number of neighbours exactly constant
by defining $h_i$ to be the distance to the $N_{\rm s}$-nearest
particle. We employ such a scheme in the serial code. Here we carry
out a range-search with $R=1.2\tilde{h}_i$, on average resulting in
$\sim 2N_{\rm s}$ potential neighbours. From these we select the
closest $N_{\rm s}$ \citep[fast algorithms for doing this exist,
see][]{Pr95}. If there are fewer than $N_{\rm s}$ particles in the
search range, or if the distance of the $N_{\rm s}$-nearest particle
inside the search range is larger than $R$, the search is repeated for
a larger search range.  In the first timestep no previous $h_i$ is
known, so we follow the neighbour tree backwards from the leaf of the
particle under consideration, until we obtain a first reasonable guess
for the local particle density (based on the number $N$ of particles
in a node of volume $l^3$), providing an initial guess for
$\tilde{h}_i$.

However, the above scheme for keeping the number of neighbours exactly
fixed is not easily accommodated in our parallel SPH implementation,
because SPH particles may have a search radius that overlaps with
several processor domains. In this case, the selection of the closest
$N_{\rm s}$ neighbours becomes non-trivial, because the underlying
data is distributed across several independent processor elements.
For parallel SPH, we therefore revert to the simpler scheme and allow
the number of neighbours to fluctuate within a small band.

\section{Time integration} \label{sectime}

As a time integrator, we use a variant of the leapfrog involving an
explicit prediction step. The latter is introduced to accommodate
individual particle timesteps in the N-body scheme, as explained later
on.

We start by describing the integrator for a single particle.
First, a particle position at the middle of the timestep $\Delta t$ is
predicted according to 
\be \tilde\vec{r}^{(n+\frac{1}{2})} =
\vec{r}^{(n)} + \vec{v}^{(n)} \frac{\Delta t}{2} , 
\ee and an
acceleration based on this position is computed, viz.  
\be
\vec{a}^{({n}+\frac{1}{2})} = -\left.\nabla\Phi
\right|_{\tilde\vec{r}^{({n}+\frac{1}{2})}}.  
\ee 
Then the particle is
advanced according to 
\be \vec{v}^{(n+1)}=\vec{v}^{(n)} +
\vec{a}^{({n}+\frac{1}{2})} \Delta t, 
\ee 
\be \vec{r}^{({n}+1)} =
\vec{r}^{({n})} + \frac{1}{2} \left[ \vec{v}^{({n})} +
\vec{v}^{({n}+1)} \right] \Delta t.  
\ee

\subsection{Timestep criterion\label{secticrit}}

In the above scheme, the timestep may vary from step to step. It is
clear that the choice of timestep size is very important in
determining the overall accuracy and computational efficiency of the
integration.

In a static potential $\Phi$, the error in specific energy arising in
one step with the above integrator is
\bea
\Delta E 
&
= 
& 
\frac{1}{4}  
\frac{\partial^2\Phi}{\partial x_i \partial x_j}
v_i^{(n)} a^{(n+\frac{1}{2})}_j \Delta t^3  
+  
\label{interr} \\
& &
\frac{1}{24}
\frac{\partial^3\Phi}{\partial x_i \partial x_j \partial x_k }
v^{(n)}_i v^{(n)}_j v^{(n)}_k  \Delta t^3
+{\cal O}(\Delta t^4) \nonumber
\eea
to leading order in $\Delta t$, i.e.\ the integrator is second order
accurate.  Here the derivatives of the potential are taken at
coordinate $\vec{r}^{(n)}$ and summation over repeated coordinate
indices is understood.

In principle, one could try to use equation (\ref{interr}) directly to
obtain a timestep by imposing some upper limit on the tolerable error
$\Delta E$.  However, this approach is quite subtle in practice.
First, the derivatives of the potential are difficult to obtain, and
second, there is no explicit guarantee that the terms of higher order
in $\Delta t$ are really small.

High-order Hermite schemes use timestep criteria that include the
first and second time derivative of the acceleration
\citep[e.g.][]{Makino91,Makino92}. While these timestep criteria are
highly successful for the integration of very nonlinear systems, they
are probably not appropriate for our low-order scheme, apart from the
fact that substantial computational effort is required to evaluate
these quantities directly. Ideally, we therefore want to use a
timestep criterion that is only based on dynamical quantities that are
either already at hand or are relatively cheap to compute.

Note that a well known problem of adaptive timestep schemes is that
they will usually break the time reversibility and symplectic nature
of the simple leapfrog. As a result, the system does not evolve under
a pseudo-Hamiltonian any more and secular drifts in the total energy
can occur. As \citet{Qu97} show, reversibility can be obtained with a
timestep that depends only on the relative coordinates of
particles. This is for example the case for timesteps that depend only
on acceleration or on local density.  However, to achieve
reversibility the timestep needs to be chosen based on the state of
the system in the middle of the timestep \citep{Qu97}, or on the
beginning {\em and} end of the timestep \citep{Hut95}. In practice,
this can be accomplished by discarding trial timesteps
appropriately. The present code selects the timestep based on the
previous step and is thus not reversible in this way.

One possible timestep criterion is obtained by constraining the
absolute size of the second order displacement of the kinetic energy,
assuming a typical velocity dispersion $\sigma^2$ for the particles,
which corresponds to a scale $E=\sigma^2$ for the typical specific
energy.  This results in \be \Delta t =\alpha_{\rm tol}
\frac{\sigma}{|\vec{a}|}
\label{eqtc2}.
\ee For a collisionless fluid, the velocity scale $\sigma$ should
ideally be chosen as the {\em local} velocity dispersion, leading to
smaller timesteps in smaller haloes, or more generally, in `colder'
parts of the fluid. The local velocity dispersion can be estimated
from a local neighbourhood of particles, obtained as in the normal SPH
formalism.

Alternatively, one can constrain the second order term in the particle
displacement, obtaining \be \Delta t = \sqrt{ \frac{ 2 \alpha'_{\rm
tol} \epsilon}{ |\vec{a}|}}
\label{eqtc3}. 
\ee Here some length scale $\alpha'_{\rm tol} \epsilon$ is introduced,
which will typically be related to the gravitational softening. This
form has quite often been employed in cosmological simulations,
sometimes with an additional restriction on the displacement of
particles in the form $\Delta t = \tilde\alpha \epsilon / |\vec{v}|$.
It is unclear though why the timesteps should depend on the
gravitational softening length in this way. In a well-resolved halo,
most orbits are not expected to change much if the halo is modeled
with more particles and a correspondingly smaller softening length, so
it should not be necessary to increase the accuracy of the time
integration for all particles by the same factor if the mass/length
resolution is increased.

For self-gravitating collisionless fluids, another plausible timestep
criterion is based on the local dynamical time: \be \Delta t =
\alpha''_{\rm tol} \frac{3}{\sqrt{8\pi G \rho}}
\label{eqtc4}. 
\ee One advantage of this criterion is that it provides a
monotonically decreasing timestep towards the center of a halo. On the
other hand, it requires an accurate estimate of the local density,
which may be difficult to obtain, especially in regions of low
density.  In particular, \citet{Qu97} have shown that haloes in
cosmological simulations that contain only a small number of
particles, about equal or less than the number employed to estimate
the local density, are susceptible to destruction if a timestep based
on (\ref{eqtc4}) is used. This is because the kernel estimates of the
density are too small in this situation, leading to excessively long
timesteps in these haloes.

In simple test integrations of singular potentials, we have found the
criterion (\ref{eqtc2}) to give better results compared to the
alternative (\ref{eqtc3}). However, neither of these simple criteria
is free of problems in typical applications to structure formation, as
we will later show in some test calculations. In the center of haloes,
subtle secular effects can occur under conditions of coarse
integration settings. The criterion based on the dynamical time does
better in this respect, but it does not work well in regions of very
low density. We thus suggest to use a combination of (\ref{eqtc2}) and
(\ref{eqtc4}) by taking the minimum of the two timesteps. This
provides good integration accuracy in low density environments and
simultaneously does well in the central regions of large haloes. For
the relative setting of the dimensionless tolerance parameters we use
$\alpha''\simeq 3 \alpha$, which typically results in a situation
where roughly the same number of particles are constrained by each of
the two criteria in an evolved cosmological simulation.  The combined
criterion is Galilean-invariant and does not make an explicit
reference to the gravitational softening length employed.

\subsection{Integrator for N-body systems}

In the context of stellar dynamical integrations, individual particle
timesteps have long been used since they were first introduced by
\citet{Aar63}. We here employ an integrator with completely flexible
timesteps, similar to the one used by \citet{Gr97} and \citet{Hi91}.
This scheme differs slightly from more commonly employed binary
hierarchies of timesteps \citep[e.g.][]{He89,McMill93,St96}.

Each particle has a timestep $\Delta t_i$, and a
current time $t_i$, where its dynamical state ($\vec{r}_i$,
$\vec{v}_i$, $\vec{a}_i$) is stored.  The dynamical state of the
particle can be predicted at times $t\in[t_i\pm0.5\Delta t_i]$ with
first order accuracy.

The next particle $k$ to be advanced is then the one with the minimum
prediction time defined as $\tau_{\rm p} \equiv {\rm min}\left( t_i +
0.5\Delta t_i \right)$.  The time $\tau_{\rm p}$ becomes the new
current time of the system.  To advance the corresponding particle, we
first predict positions for {\em all} particles at time $\tau_{\rm p}$
according to
\be
\tilde\vec{r}_{i} = 
\vec{r}_i + \vec{v}_i ( \tau_{\rm p}- t_i ).
\ee
Based on these positions, the acceleration of particle $k$ at the
middle of its timestep is calculated as
\be
\vec{a}_k^{({n}+\frac{1}{2})} = -\left.\nabla\Phi(\tilde\vec{r}_{i})
\right|_{\tilde\vec{r}_k} .
\ee
Position and velocity of particle $k$ are then advanced as
\be
\vec{v}^{(n+1)}_k=\vec{v}^{(n)}_k + 2 \vec{a}_k^{({n}+\frac{1}{2})}  ( \tau_{\rm p}- t_k),
\ee
\be
\label{eq44}
\vec{r}^{(n+1)}_k = 
\vec{r}^{(n)}_k 
+ \left[ \vec{v}^{(n)}_k + \vec{v}^{({n}+1)}_k \right] 
 (\tau_{\rm p}- t_k)  ,
\ee
and its current time can be updated to
\be
t_k^{(\rm new)} = t_k +  2( \tau_{\rm p}- t_k).
\ee
Finally, a new timestep $\Delta t_k^{(\rm new)}$ for the particle is
estimated.

At the beginning of the simulation, all particles start out with the
same current time.  However, since the timesteps of the particles are
all different, the current times of the particles distribute
themselves nearly symmetrically around the current prediction time,
hence the prediction step involves forward and backward prediction to
a similar extent.

Of course, it is impractical to advance only a single particle at any
given prediction time, because the prediction itself and the (dynamic)
tree updates induce some overhead.  For this reason we advance
particles in bunches.  The particles may be thought of as being
ordered according to their prediction times $t_i^{\rm p}= t_i +
\frac{1}{2} \Delta t_i$. The simulation works through this time line,
and always advances the particle with the smallest $t_i^{\rm p}$, and
also all subsequent particles in the time line, until the first is
found with \be \tau_{\rm p} \le t_i + \frac{1}{4} \Delta t_i \, .  \ee
This condition selects a group of particles at the lower end of the
time line, and all the particles of the group are guaranteed to be
advanced by at least half of their maximum allowed timestep. Compared
to using a fixed block step scheme with a binary hierarchy, particles
are on average advanced closer to their maximum allowed timestep in
this scheme, which results in a slight improvement in
efficiency. Also, timesteps can more gradually vary than in a power of
two hierarchy. However, a perhaps more important advantage of this
scheme is that it makes work-load balancing in the parallel code
simpler, as we will discuss in more detail later on.

In practice, the size $M$ of the group that is advanced at a given
step is often only a small fraction of the total particle number
$N$. In this situation it becomes important to eliminate any overhead
that scales with ${\cal O}(N)$. For example, we obviously need to find
the particle with the minimum prediction time at every timestep, and
also the particles following it in the time line. A loop over all the
particles, or a complete sort at every timestep, would induce overhead
of order ${\cal O}(N)$ or ${\cal O}(N\log N)$, which can become
comparable to the force computation itself if $M/N \ll 1$.  We solve
this problem by keeping the maximum prediction times of the particles
in an ordered binary tree \citep{Wi86} at all times.  Finding the
particle with the minimum prediction time and the ones that follow it
are then operations of order ${\cal O}(\log N)$. Also, once the
particles have been advanced, they can be removed and reinserted into
this tree with a cost of order ${\cal O}(\log N)$. Together with the
dynamic tree updates, which eliminate prediction and tree construction
overhead, the cost of the timestep then scales as ${\cal O}(M\log N)$.

\subsection{Dynamic tree updates}

If the fraction of particles to be advanced at a given timestep is
indeed small, the prediction of {\em all} particles and the
reconstruction of the {\em full} tree would also lead to significant
sources of overhead.  However, as \citet{McMill93} have first
discussed, the geometric structure of the tree, i.e.\ the way the
particles are grouped into a hierarchy, evolves only relatively slowly
in time. It is therefore sufficient to reconstruct this grouping only
every few timesteps, provided one can still obtain accurate node
properties (center of mass, multipole moments) at the current
prediction time.

We use such a scheme of dynamic tree updates by predicting properties
of tree-nodes on the fly, instead of predicting all particles every
single timestep. In order to do this, each node carries a
center-of-mass velocity in addition to its position at the time of its
construction. New node positions can then be predicted while the tree
is walked, and only nodes that are actually visited need to be
predicted.  Note that the leaves of the tree point to single
particles. If they are used in the force computation, their prediction
corresponds to the ordinary prediction as outlined in equation
(\ref{eq44}).

In our simple scheme we neglect a possible time variation of the
quadrupole moment of the nodes, which can be taken into account in
principle \citep{McMill93}. However, we introduce a mechanism that
reacts to fast time variations of tree nodes. Whenever the center-of
mass of a tree node under consideration has moved by more than a small
fraction of the nodes' side-length since the last reconstruction of
this part of the tree, the node is completely updated, i.e.~the
center-of-mass, center-of-mass velocity and quadrupole moment are
recomputed from the individual (predicted) phase-space variables of
the particles. We also adjust the side-length of the tree node if any
of its particles should have left its original cubical volume.

Finally, the full tree is reconstructed from scratch every once in a
while to take into account the slow changes in the grouping
hierarchy. Typically, we update the tree whenever a total of $\sim
0.1N$ force computations have been done since the last full
reconstruction. With this criterion the tree construction is an
unimportant fraction of the total computation time. We have not
noticed any significant loss of force accuracy induced by this
procedure.

In summary, the algorithms described above result in an integration
scheme that can smoothly and efficiently evolve an N-body system
containing a large dynamic range in time scales. At a given timestep,
only a small number $M$ of particles are then advanced, and the total
time required for that scales as ${\cal O} (M\log N)$.

\subsection{Including SPH}

The above time integration scheme may easily be extended to include
SPH. Here we also need to integrate the internal energy equation, and
the particle accelerations also receive a hydrodynamical component.
To compute the latter we also need predicted velocities
\be
\tilde\vec{v}_{i} = 
\vec{v}_i + \vec{a}_{i-1} ( \tau_{\rm p}- t_i ) ,
\ee
where we have approximated $\vec{a}_i$ with the acceleration of the
previous timestep.  Similarly, we obtain predictions for the internal
energy
\be
\tilde{u}_{i} = 
{u}_i + \dot{u}_i ( \tau_{\rm p}- t_i ) ,
\ee
and the density of inactive particles as
\be
\tilde{\rho}_{i} = 
{\rho}_i + \dot{\rho}_i ( \tau_{\rm p}- t_i ).
\ee

For those particles that are to be advanced at the current system
step, these predicted quantities are then used to compute the
hydrodynamical part of the acceleration and the rate of change of 
internal energy with the usual SPH estimates, as described in Section
\ref{secsph}.

For hydrodynamical stability, the collisionless timestep criterion
needs to be supplemented with the Courant condition. We adopt it for
the gas particles in the form \be \Delta t _i = \frac { \alpha_{\rm
cour}\,h_i}{ h_i |(\nabla\cdot \vec{v})_i| + \max(c_i, |\vec{v}_i|) \,
(1+0.6\,\alpha_{\rm visc}) },
\label{eqtc5}
\ee where $\alpha_{\rm visc}$ regulates the strength of the artificial
bulk viscosity, and $\alpha_{\rm cour}$ is an accuracy parameter, the
Courant factor. Note that we use the maximum of the sound speed $c_i$
and the bulk velocity $|\vec{v}_i|$ in this expression. This improves
the handling of strong shocks when the infalling material is cold, but
has the disadvantage of not being Galilean invariant.  For the
SPH-particles, we use either the adopted criterion for collisionless
particles or (\ref{eqtc5}), whichever gives the smaller timestep.

As defined above, we evaluate $\vec{a}^{\rm gas}$ and $\dot{u}$ at the
middle of the timestep, when the actual timestep $\Delta t$ of the
particle that will be advanced {\em is already set}.  Note that there
is a term in the artificial viscosity that can cause a problem in this
explicit integration scheme.  The second term in equation
(\ref{eqvisc}) tries to prevent particle inter-penetration.  If a
particle happens to get relatively close to another SPH particle in
the time $\Delta t/2$ and the relative velocity of the approach is
large, this term can suddenly lead to a very large repulsive
acceleration $\vec{a}^{\rm visc}$, trying to prevent the particles
from getting any closer. However, it is then too late to reduce the
timestep. Instead, the velocity of the approaching particle will be
changed by $\vec{a}^{\rm visc} \Delta t$, possibly {\em reversing} the
approach of the two particles. But the artificial viscosity should at
most {\em halt} the approach of the particles.  To guarantee this, we
introduce an upper cut-off to the maximum acceleration induced by the
artificial viscosity.  If $\vec{v}_{ij}\cdot\vec{r}_{ij}<0$, we
replace equation (\ref{eqvisc0}) with \be
\tilde\Pi_{ij}=\frac{1}{2}(f_i+f_j)\,\min\left[\Pi_{ij},
\frac{\vec{v}_{ij}\cdot\vec{r}_{ij}} {(m_i+m_j) W_{ij} \Delta t}
\right], \ee where $W_{ij}= \vec{r}_{ij} \nabla_i \left[
W(\vec{r}_{ij};h_i)+W(\vec{r}_{ij};h_j)\right]/2$.  With this change,
the integration scheme still works reasonably well in regimes with
strong shocks under conditions of relatively coarse timestepping. Of
course, a small enough value of the Courant factor will prevent this
situation from occurring to begin with.

Since we use the gravitational tree of the SPH particles for the
neighbour search, another subtlety arises in the context of dynamic
tree updates, where the full tree is not necessarily reconstructed
every single timestep. The range searching technique relies on the
current values of the maximum SPH smoothing length in each node, and
also expects that all particles of a node are still inside the
boundaries set by the side-length of a node. To guarantee that the
neighbour search will always give correct results, we perform a
special update of the SPH-tree every timestep. It involves a loop over
every SPH particle that checks whether the particle's smoothing length
is larger than $h_{\rm max}$ stored in its parent node, or if it falls
outside the extension of the parent node. If either of these is the
case, the properties of the parent node are updated accordingly, and
the tree is further followed `backwards' along the parent nodes, until
each node is again fully `contained' in its parent node. While this
update routine is very fast in general, it does add some overhead,
proportional to the number of SPH particles, and thus breaks in
principle the ideal scaling (proportional to $M$) obtained for purely
collisionless simulations.

\subsection{Implementation of cooling}

When radiative cooling is included in simulations of galaxy formation
or galaxy interaction, additional numerical problems arise. In regions
of strong gas cooling, the cooling times can become so short that
extremely small timesteps would be required to follow the internal
energy accurately with the simple explicit integration scheme used so
far.

To remedy this problem, we treat the cooling semi-implicitly in an
isochoric approximation. At any given timestep, we first compute the
rate $\dot{u}^{\rm ad}$ of change of the internal energy due to the
ordinary adiabatic gas physics.  In an isochoric approximation, we
then solve implicitly for a new internal energy predicted at the end
of the timestep, i.e.  \be \hat{u}_i^{(n+1)} = u_i^{(n)} +
\dot{u}^{\rm ad}\Delta t - \frac{\Lambda\left[\rho_i^{(n)},
\hat{u}_i^{(n+1)}\right]\Delta t}{\rho_i^{(n)}} \, .  \ee The implicit
computation of the cooling rate guarantees stability.  Based on this
estimate, we compute an effective rate of change of the internal
energy, which we then take as \be \dot{u}_i=\left[\hat{u}_i^{(n+1)} -
u_i^{(n)}\right]/\Delta t \,.  \ee We use this last step because the
integration scheme requires the possibility to predict the internal
energy at arbitrary times. With the above procedure, $u_i$ is always a
continuous function of time, and the prediction of $u_i$ may be done
for times in between the application of the isochoric
cooling/heating. Still, there can be a problem with predicted internal
energies in cases when the cooling time is very small. Then a particle
can lose a large fraction of its energy in a single timestep. While
the implicit solution will still give a correct result for the
temperature at the end of the timestep, the predicted energy in the
middle of the {\em next} timestep could then become very small or even
negative because of the large negative value of $\dot u$. We therefore
restrict the maximum cooling rate such that a particle is only allowed
to lose at most half its internal energy in a given timestep,
preventing the predicted energies from `overshooting'. \citet{Ka91}
have used a similar method to damp the cooling rate.

\subsection{Integration in comoving coordinates}

For simulations in a cosmological context, the expansion of the
universe has to be taken into account. Let $\vec{x}$ denote comoving
coordinates, and $a$ be the dimensionless scale factor ($a=1.0$ at the
present epoch).  Then the Newtonian equation of motion becomes
\be
\label{eq52}
\ddot{\vec{x}} +2\frac{\dot{a}}{a}\,\dot{\vec{x}} =-G\int
\frac{\delta\rho(\vec{x}')\,(\vec{x}-\vec{x}')}{|\vec{x}-\vec{x}'|^3}
\dd^3 x'.
\ee
Here the function $\delta \rho(\vec{x}) = \rho(\vec{x})- \ol{\rho}$
denotes the (proper) density fluctuation field.

In an N-body simulation with periodic boundary conditions, the volume
integral of equation (\ref{eq52}) is carried out over {\em all}
space. As a consequence, the homogeneous contribution arising from
$\ol{\rho}$ drops out around every point. Then the equation of motion
of particle $i$ becomes 
\be \ddot{\vec{x}}_i
+2\frac{\dot{a}}{a}\,\dot{\vec{x}}_i =-\frac{G}{a^3}\sum_{j\ne i \atop
{\rm periodic}} \frac{
m_j\,(\vec{x}_i-\vec{x}_j)}{|\vec{x}_i-\vec{x}_j|^3}, 
\ee 
where the summation includes all periodic images of the particles $j$.

However, one may also employ vacuum boundary conditions if one
simulates a spherical region of radius $R$ around the origin, and
neglects density fluctuations outside this region.  In this case, the
background density $\ol{\rho}$ gives rise to an additional term, viz.\
\be
\label{eq55}
\ddot{\vec{x}}_i +2\frac{\dot{a}}{a}\,\dot{\vec{x}}_i
=\frac{1}{a^3}\left[ -G\sum_{j\ne i } \frac{
m_j\,\vec{x}_{ij}}{|\vec{x}_{ij}|^3} \,+\,\frac{1}{2}\Omega_0 H_0^2
\vec{x}_i \right] .  
\ee 
\gadget\ supports both periodic and vacuum boundary conditions.
We implement the former by means of the Ewald summation technique
\citep{He91}.

For this purpose, we modify the tree walk such that each node is
mapped to the position of its nearest periodic image with respect to
the coordinate under consideration.  If the multipole expansion of the
node can be used according to the cell opening criterion, its partial
force is computed in the usual way. However, we also need to add the
force exerted by all the other periodic images of the node. The
slowly converging sum over these contributions can be evaluated with
the Ewald technique.  If $\vec{x}$ is the coordinate of the point of
force-evaluation relative to a node of mass $M$, the resulting {\em
additional} acceleration is given by
\bea 
\vec{a}_c(\vec{x})=
M \Bigg\{\frac{\vec{x}}{|\vec{x}|^3} - \sum_\vec{n}
\frac{\vec{x}-\vec{n}L}{|\vec{x}-\vec{n}L|^3}\times \bigg[{\rm
erfc}(\alpha|\vec{x}-\vec{n}L|) + \nonumber \hspace{-1cm} \\
 \frac{2\alpha|\vec{x}-\vec{n}L|}{\sqrt\pi}\exp\left(-\alpha^2|\vec{x}-\vec{n}L|^2\right)
\bigg] \nonumber \\  
- \frac{2}{L^2}\sum_{\vec{h}\ne 0}
\frac{\vec{h}}{|\vec{h}|^2} \exp\left(-\frac{\pi^2
|\vec{h}|^2}{\alpha^2 L^2}\right)
\sin\left(\frac{2\pi}{L}\vec{h}\cdot\vec{x}\right) \Bigg\}.  
\eea
Here $\vec{n}$ and $\vec{h}$ are integer triplets, $L$ is the box
size, and $\alpha$ is an arbitrary number \citep{He91}.  Good
convergence is achieved for $\alpha=2/L$, where we sum over the
range $|\vec{n}|<5$ and $|\vec{h}|<5$.  Similarly, the additional 
potential due to the periodic replications of the node is given by 
\bea
\phi_c(\vec{x})= M\bigg\{ \frac{1}{\vec{x}} + \frac{\pi}{\alpha^2 L^3}
- \sum_{\vec{n}} \frac{{\rm
erfc}(\alpha|\vec{x}-\vec{n}L|)}{|\vec{x}-\vec{n}L|} -
\nonumber\\
\frac{1}{L}\sum_{\vec{h}\ne 0}  \frac{1}{\pi |\vec{h}|^2}\exp\left(-\frac{\pi^2 |\vec{h}|^2}{\alpha^2 L^2}\right)
\cos\left(\frac{2\pi}{L}\vec{h}\cdot\vec{x}\right) \Bigg\}.
\eea

We follow \cite{He91} and tabulate the correction fields
$\vec{a}_c(\vec{x})/M$ and $\phi_c(\vec{x})/M$ for one octant of the
simulation box, and obtain the result of the Ewald summation during
the tree walk from trilinear interpolation off this grid.  It should
be noted however, that periodic boundaries have a strong impact on the
speed of the tree algorithm. The number of floating point operations
required to interpolate the correction forces from the grid has a
significant toll on the raw force speed and can slow it down by almost
a factor of two.

In linear theory, it can be shown that the kinetic energy \be T=
\frac{1}{2} \sum_i m_i {\vec{v}}^2 \ee in peculiar motion grows
proportional to $a$, at least at early times. This implies that
$\sum_i m_i \dot{\vec{x}}^2 \propto 1/a$, hence the comoving
velocities $\dot{\vec{x}}=\vec{v}/a$ actually diverge for $a\to 0$.
Since cosmological simulations are usually started at redshift
$z\simeq 30-100$, one therefore needs to follow a rapid deceleration
of $\dot{\vec{x}}$ at high redshift. So it is numerically unfavourable
to solve the equations of motion in the variable $\dot{\vec{x}}$.

To remedy this problem, we use an alternative velocity variable
\be
\vec{w}\equiv a^{\frac{1}{2}} \,\dot{\vec{x}},
\label{inq1}
\ee
and we employ the expansion factor itself as time variable.  Then the
equations of motion become
\be
\frac{\dd \vec{w}}{\dd a}
=-\frac{3}{2} \frac{\vec{w}}{a} +\frac{1}{a^2 S(a)}
\left[
-G\sum_{j\ne i } \frac{ m_j\,\vec{x}_{ij}}{|\vec{x}_{ij}|^3}
\,+\,\frac{1}{2}\Omega_0 H_0^2 \vec{x}_i
\right],
\label{cosmo0}
\ee
\be
\frac{\dd \vec{x}}{\dd a} = \frac{\vec{w}}{S(a)},
\ee
with $S(a)= a^{\frac{3}{2}} H(a)$ given by
\be
S(a)=H_0 \sqrt{ \Omega_0
+a(1-\Omega_0-\Omega_{\Lambda})+a^3\Omega_\Lambda}.
\label{inq3}
\ee 
Note that for periodic boundaries the second term in the square
bracket of equation (\ref{cosmo0}) is absent, instead the summation
extends over all periodic images of the particles.

Using the Zel'dovich approximation, one sees that $\vec{w}$ remains
constant in the linear regime.  Strictly speaking this holds at all
times only for an Einstein-de-Sitter universe, however, it is also
true for other cosmologies at early times.  Hence equations
(\ref{inq1}) to (\ref{inq3}) in principle solve linear theory for
arbitrarily large steps in $a$. This allows to traverse the linear
regime with maximum computational efficiency.  Furthermore, equations
(\ref{inq1}) to (\ref{inq3}) represent a convenient formulation for
general cosmologies, and for our variable timestep integrator. Since
$\vec{w}$ does not vary in the linear regime, predicted particle
positions based on $\tilde\vec{x}_i= \vec{x}_i +\vec{w}_i(a_{\rm
p}-a_i)/S(a_{\rm p})$ are quite accurate. Also, the acceleration
entering the timestep criterion may now be identified with $\dd\vec{w}
/\dd a$, and the timestep (\ref{eqtc2}) becomes \be \Delta a =
\alpha_{\rm tol}{\sigma} {\left| \frac{\dd \vec{w}}{\dd
a}\right|^{-1}}.  \ee

The above equations only treat the gravity part of the dynamical
equations. However, it is straightforward to express the
hydrodynamical equations in the variables ($\vec{x}$, $\vec{w}$, $a$)
as well.  For gas particles, equation (\ref{cosmo0}) receives an
additional contribution due to hydrodynamical forces, viz.  
\be \left( \frac{\dd \vec{w}}{\dd a}\right)_{\rm hydro} =
-\frac{1}{a\,S(a)}\,\frac{\nabla_{\vec{x}}P}{\rho}.  \ee For the
energy equation, one obtains \be \frac{\dd u}{\dd a}=
-\frac{3}{a}\,\frac{P}{\rho} - \frac{1}{S(a)}\,\frac{P}{\rho}\,
\nabla_{\vec{x}}\cdot\vec{w}.  
\ee
Here the first term on the right hand side describes the adiabatic
cooling of gas due to the expansion of the universe.

\begin{figure*}
\bc
\resizebox{13cm}{!}
{\includegraphics{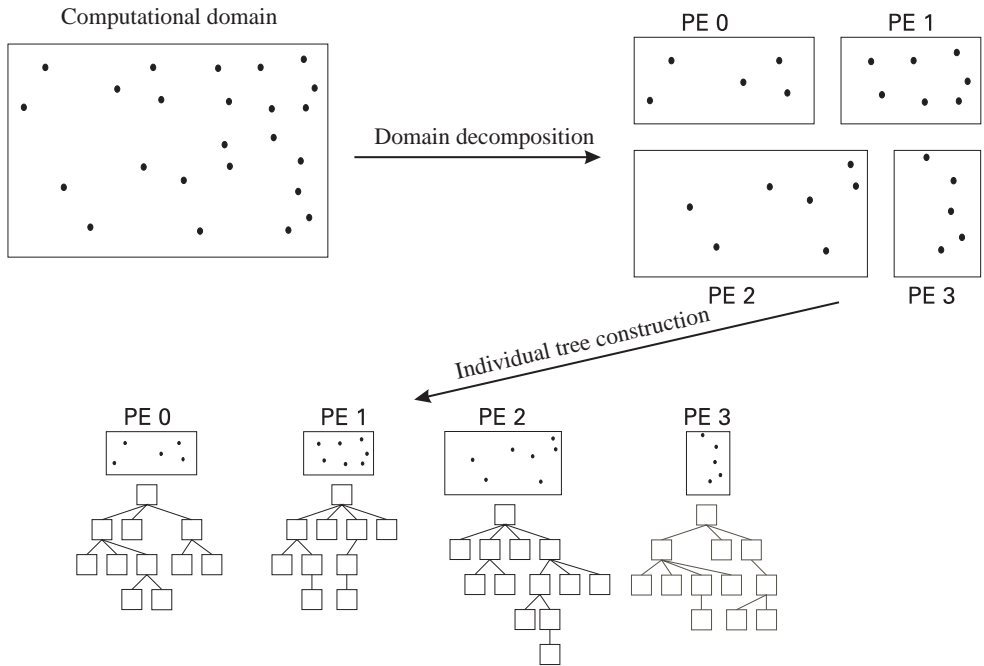}}\\
\caption{Schematic representation of the domain decomposition in two
dimensions, and for four processors. Here, the first split occurs
along the y-axis, separating the processors into two groups.  They
then independently carry out a second split along the x-axis. After
completion of the domain decomposition, each processor element (PE)
can construct its own BH tree just for the particles in its part of
the computational domain.
\label{figdomain}}
\ec
\end{figure*}

\section{Parallelization} \label{secpara} 

Massively parallel computer systems with distributed memory have
become increasingly popular recently. They can be thought of as a
collection of workstations, connected by a fast communication
network. This architecture promises large scalability for reasonable
cost. Current state-of-the art machines of this type include the Cray
T3E and IBM SP/2. It is an interesting development that `Beowulf'-type
systems based on commodity hardware have started to offer floating
point performance comparable to these supercomputers, but at a much
lower price.

However, an efficient use of parallel distributed memory machines
often requires substantial changes of existing algorithms, or the
development of completely new ones. Conceptually, parallel
programming involves two major difficulties in addition to the task of
solving the numerical problem in a serial code. First, there is the
difficulty of how to divide the work and data {\em evenly} among the
processors, and second, an efficient communication scheme between the
processors needs to be devised.

In recent years, a number of groups have developed parallel N-body
codes, all of them with different parallelization strategies, and
different strengths and weaknesses.  Early versions of parallel codes
include those of \citet{Barnes86}, \citet{MH89} and
\citet{Th93}. Later, \citet{Wa92} parallelized the BH-tree code on
massively parallel machines with distributed memory. \citet{Du96b}
presented the first parallel tree code based on MPI. \citet{Di95} have
developed a parallel simulation code ({\small PKDGRAV}) that works
with a balanced binary tree.  More recently, parallel tree-SPH codes
have been introduced by \citet{Da97} and \citet{Lia99}, and a PVM
implementation of a gravity-only tree code has been described by
\citet{Vit2000}.

We here report on our newly developed parallel version of \gadget,
where we use a parallelization strategy that differs from previous
workers. It also implements individual particle timesteps for the
first time on distributed-memory, massively parallel computers.  We
have used the {\em Message Passing Interface} (MPI) \citep{Sn95,Pa97},
which is an explicit communication scheme, i.e.\ it is entirely up to
the user to control the communication. Messages containing data can be
sent between processors, both in synchronous and asynchronous modes.
A particular advantage of MPI is its flexibility and portability. Our
simulation code uses only standard C and standard MPI, and should
therefore run on a variety of platforms. We have confirmed this so far
on Cray T3E and IBM SP/2 systems, and on Linux-PC clusters.

\subsection{Domain decomposition}

The typical size of problems attacked on parallel computers is usually
much too large to fit into the memory of individual computational
nodes, or into ordinary workstations.  This fact alone (but of course
also the desire to distribute the work among the processors) requires
a partitioning of the problem onto the individual processors.

\begin{figure*}
\bc
\resizebox{13cm}{!}
{\includegraphics{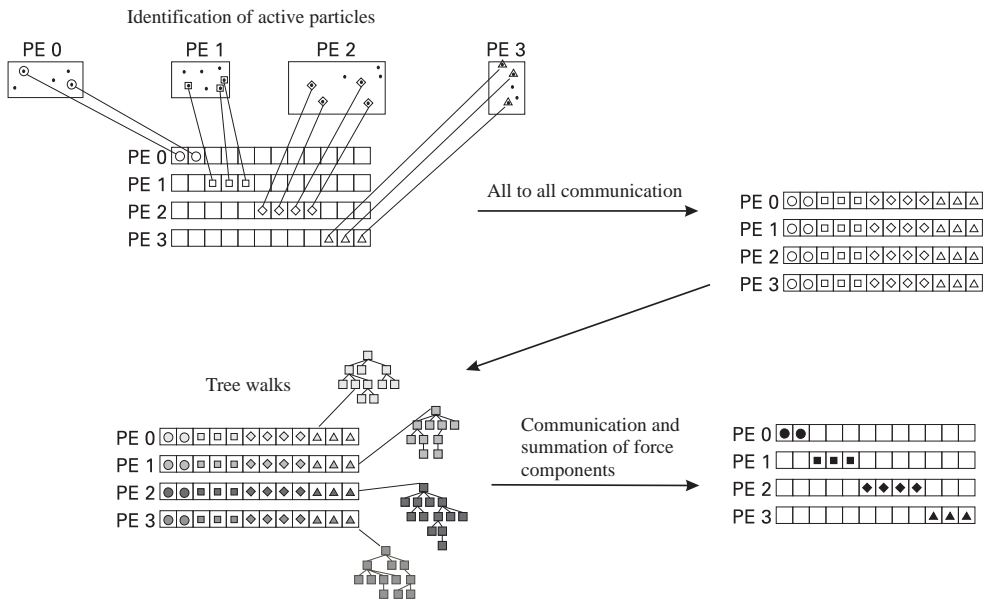}}\\
\caption{Schematic illustration of the parallelization scheme of
\gadget\ for the force computation. In the first step, each PE
identifies the active particles, and puts their coordinates in a
communication buffer. In a communication phase, a single and identical
list of all these coordinates is then established on all processors.
Then each PE walks its local tree for this list, thereby obtaining a
list of partial forces. These are then communicated in a collective
process back to the original PE that hosts the corresponding particle
coordinate. Each processor then sums up the incoming force
contributions, and finally arrives at the required total forces for
its active particles.
\label{figparallel}}
\ec
\end{figure*}

For our N-body/SPH code we have implemented a spatial domain
decomposition, using the orthogonal recursive bisection (ORB)
algorithm \citep{Du96b}.  In the first step, a split is found along
one spatial direction, e.g.\ the x-axis, and the collection of
processors is grouped into two halves, one for each side of the
split. These processors then exchange particles such that they end up
hosting only particles lying on their side of the split.  In the
simplest possible approach, the position of the split is chosen such
that there are an equal number of particles on both sides. However,
for an efficient simulation code the split should try to balance the
work done in the force computation on the two sides.  This aspect will
be discussed further below.

In a second step, each group of processors finds a new split along a
different spatial axis, e.g.\ the y-axis. This splitting process is
repeated recursively until the final groups consist of just one
processor, which then hosts a rectangular piece of the computational
volume.  Note that this algorithm constrains the number of processors
that may be used to a power of two. Other algorithms for the domain
decomposition, for example Voronoi tessellations \citep{Yah99}, are
free of this restriction.

A two-dimensional schematic illustration of the ORB is shown in Figure
\ref{figdomain}. Each processor can construct a local BH tree for its
domain, and this tree may be used to compute the force exerted by the
processors' particles on arbitrary test particles in space.

\subsection{Parallel computation of the gravitational force}

\gadget's algorithm for parallel force computation differs from that
of \citet{Du96b}, who introduced the notion of {\em locally essential
trees}. These are trees that are sufficiently detailed to allow the
full force computation for any particle local to a processor, without
further need for information from other processors. The locally
essential trees can be constructed from the local trees by pruning and
exporting parts of these trees to other processors, and attaching
these parts as new branches to the local trees. To determine which
parts of the trees need to be exported, special tree walks are
required.

A difficulty with this technique occurs in the context of dynamic tree
updates. While the additional time required to promote local trees to
locally essential trees should not be an issue for an integration
scheme with a global timestep, it can become a significant source of
overhead in individual timestep schemes. Here, often only one per cent
or less of all particles require a force update at one of the (small)
system timesteps. Even if a dynamic tree update scheme is used to
avoid having to reconstruct the full tree every timestep, the locally
essential trees are still confronted with subtle synchronization
issues for the nodes and particles that have been imported from other
processor domains. Imported particles in particular may have received
force computations since the last `full' reconstruction of the locally
essential tree occurred, and hence need to be re-imported. The local
domain will also lack sufficient information to be able to update
imported nodes on its own if this is needed. So some additional
communication needs to occur to properly synchronize the locally
essential trees on each timestep. `On-demand' schemes, involving
asynchronous communication, may be the best way to accomplish this in
practice, but they will still add some overhead and are probably quite
complicated to implement.  Also note that the construction of locally
essential trees depends on the opening criterion.  If the latter is
not purely geometric but depends on the particle for which the force
is desired, it can be difficult to generate a fully sufficient locally
essential tree.  For these reasons we chose a different
parallelization scheme that scales linearly with the number of
particles that need a force computation.

Our strategy starts from the observation that each of the local
processor trees is able to provide the force exerted by its particles
for any location in space. The full force might thus be obtained by
adding up all the partial forces from the local trees. As long as the
number of these trees is less than the number of typical particle-node
interactions, this computational scheme is practically not more
expensive than a tree walk of the corresponding locally essential
tree.

A force computation therefore requires a communication of the desired
coordinates to all processors. These then walk their local trees, and
send partial forces back to the original processor that sent
out the corresponding coordinate.  The total force is then obtained by
summing up the incoming contributions.

In practice, a force computation for a {\em single} particle would be
badly imbalanced in work in such a scheme, since some of the
processors could stop their tree walk already at the root node, while
others would have to evaluate several hundred particle-node
interactions. However, the time integration scheme advances at a given
timestep always a group of $M$ particles of size about 0.5-5 per cent
of the total number of particles. This group represents a
representative mix of the various clustering states of matter in the
simulation.  Each processor contributes some of its particle positions
to this mix, but the total list of coordinates is the same for all
processors. If the domain decomposition is done well, one can arrange
that the cumulative time to walk the local tree for all coordinates in
this list is the same for all processors, resulting in good work-load
balance. In the time integration scheme outlined above, the size $M$
of the group of {\em active} particles is always roughly the same from
step to step, and it also represents always the same statistical mix
of particles and work requirements. This means that the same domain
decomposition is appropriate for each of a series of consecutive
steps. On the other hand, in a block step scheme with binary
hierarchy, a step where all particles are synchronized may be followed
by a step where only a very small fraction of particles are active. In
general, one cannot expect that the same domain decomposition will
balance the work for both of these steps.

Our force computation scheme proceeds therefore as sketched
schematically in Figure \ref{figparallel}. Each processor identifies
the particles that are to be advanced in the current timestep, and
puts their coordinates in a communication buffer. Next, an all-to-all
communication phase is used to establish the same list of coordinates
on all processors. This communication is done in a collective fashion:
For $N_{\rm p}$ processors, the communication involves $N_{\rm p}-1$
cycles. In each cycle, the processors are arranged in $N_{\rm p}/2$
pairs. Each pair exchanges their original list of active coordinates.
While the amount of data that needs to be communicated scales as
${\cal O}[M (N_{\rm p}-1)] \simeq {\cal O}(M N_{\rm p})$, the
wall-clock time required scales only as ${\cal O}(M + c\, N_{\rm p})$
because the communication is done fully in parallel. The term $c\,
N_{\rm p}$ describes losses due to message latency and overhead due to
the message envelopes.  In practice, additional losses can occur on
certain network topologies due to message collisions, or if the
particle numbers contributed to $M$ by the individual processors are
significantly different, resulting in communication imbalance. On the
T3E, the communication bandwidth is large enough that only a very
small fraction of the overall simulation time is spent in this phase,
even if processor partitions as large as 512 are used. On
Beowulf-class networks of workstations, we find that typically less
than about 10-20\% of the time is lost due to communication overhead
if the computers are connected by a switched network with a speed of
$100\,{\rm Mbit\,s^{-1}}$.

In the next step, all processors walk their local trees and replace
the coordinates with the corresponding force contribution.  Note that
this is the most time-consuming step of a collisionless simulation (as
it should be), hence work-load balancing is most crucial here. After
that, the force contributions are communicated in a similar way as
above between the processor pairs. The processor that hosted a
particular coordinate adds up the incoming force contributions and
finally ends up with the full force for that location. These forces
can then be used to advance its locally active particles, and to
determine new timesteps for them. In these phases of the N-body
algorithm, as well as in the tree construction, no further information
from other processors is required.

\subsection{Work-load balancing}

Due to the high communication bandwidth of parallel supercomputers
like the T3E or the SP/2, the time required for force computation is
dominated by the tree walks, and this is also the dominating part of
the simulation as a whole. It is therefore important that this part of
the computation parallelizes well. In the context of our
parallelization scheme, this means that the domain decomposition
should be done such that the time spent in the tree walks of each step
is the same for all processors.

It is helpful to note, that the list of coordinates for the tree walks
is {\em independent} of the domain decomposition. We can then think of
each patch of space, represented by its particles, to cause some cost
in the tree-walk process. A good measure for this cost is the number
of particle-node interactions originating from this region of space.
To balance the work, the domain decomposition should therefore try to
make this cost equal on the two sides of each domain split.

In practice, we try to reach this goal by letting each tree-node carry
a counter for the number of node-particle interactions it participated
in since the last domain decomposition occurred. Before a new domain
decomposition starts, we then assign this cost to individual particles
in order to obtain a weight factor reflecting the cost they on average
incur in the gravity computation. For this purpose, we walk the tree
backwards from a leaf (i.e.\ a single particle) to the root node. In
this walk, the particle collects its total cost by adding up its share
of the cost from all its parent nodes. The computation of these cost
factors differs somewhat from the method of \citet{Du96b}, but the
general idea of such a work-load balancing scheme is similar.

Note that an optimum work-load balance can often result in substantial
memory imbalance. Tree-codes consume plenty of memory, so that the
feasible problem size can become memory rather than CPU-time
limited. For example, a single node with 128 Mbyte on the Garching T3E
is already filled to roughly 65 per cent with 4.0$\times 10^{5}$
particles by the present code, including all memory for the tree
structures, and the time integration scheme. In this example, the
remaining free memory can already be insufficient to allow an optimum
work-load balance in strongly clustered simulations. Unfortunately,
such a situation is not untypical in practice, since one usually
strives for {\em large} $N$ in N-body work, so one is always tempted
to fill up most of the available memory with particles, without
leaving much room to balance the work-load.  Of course, \gadget\ can
only try to balance the work within the constraints set by the
available memory. The code will also strictly observe a memory limit
in all intermediate steps of the domain decomposition, because some
machines (e.g. T3E) do not have virtual memory.

\subsection{Parallelization of SPH}

Hydrodynamics can be seen as a more natural candidate for
parallelization than gravity, because it is a {\em local}
interaction. In contrast to this, the gravitational N-body problem has
the unpleasant property that at all times {\em each particle interacts
with every other particle}, making gravity intrinsically difficult to
parallelize on distributed memory machines.

It therefore comes as no large surprise that the parallelization of
SPH can be handled by invoking the same strategy we employed for the
gravitational part, with only minor adjustments that take into account
that most SPH particles (those entirely `inside' a local domain) do
not rely on information from other processors.

In particular, as in the purely gravitational case, we do not import
neighbouring particles from other processors.  Rather, we export the
particle coordinates (and other information, if needed) to other
processors, which then deliver partial hydrodynamical forces or
partial densities contributed by their particle population.  For
example, the density computation of SPH can be handled in much the
same way as that of the gravitational forces.  In a collective
communication, the processors establish a common list of all
coordinates of active particles together with their smoothing
lengths. Each processor then computes partial densities by finding
those of its particles that lie within each smoothing region, and by
doing the usual kernel weighting for them. These partial densities are
then delivered back to the processor that holds the particle
corresponding to a certain entry in the list of active
coordinates. This processor then adds up the partial contributions,
and obtains the full density estimate for the active particle.

The locality of the SPH interactions brings an important
simplification to this scheme.  If the smoothing region of a particle
lies entirely inside a local domain, the particle coordinate does not
have to be exported at all, allowing a large reduction in the length
of the common list of coordinates each processor has to work on, and
reducing the necessary amount of communication involved by a large
factor.

In the first phase of the SPH computation, we find in this way the
density and the total number of neighbours inside the smoothing
length, and we evaluate velocity divergence and curl. In the second
phase, the hydrodynamical forces are then evaluated in an analogous
fashion.

Notice that the computational cost and the communication overhead of
this scheme again scale linearly with the number $M$ of active
particles, a feature that is essential for good adaptivity of the
code. We also note that in the second phase of the SPH computation,
the particles `see' the updated density values automatically. When SPH
particles themselves are imported to form a locally essential
SPH-domain, they have to be exchanged a second time if they are active
particles themselves in order to have correctly updated density values
for the hydrodynamical force computation \citep{Da97}.

An interesting complication arises when the domain decomposition is
not repeated every timestep. In practice this means that the
boundaries of a domain may become `diffuse' when some particles start
to move out of the boundaries of the original domain.  If the spatial
region that we classified as interior of a domain is not adjusted, we
then risk missing interactions because we might not export a particle
that starts to interact with a particle that has entered the local
domain boundaries, but is still hosted by another processor. Note that
our method to update the neighbour-tree on each single timestep
already guarantees that the correct neighbouring particles are always
found on a local processor -- if such a search is conducted at all. So
in the case of soft domain boundaries we only need to make sure that
all those particles are really exported that can possibly interact
with particles on other processors.

\begin{figure}
\bc
\resizebox{8cm}{!}{\includegraphics{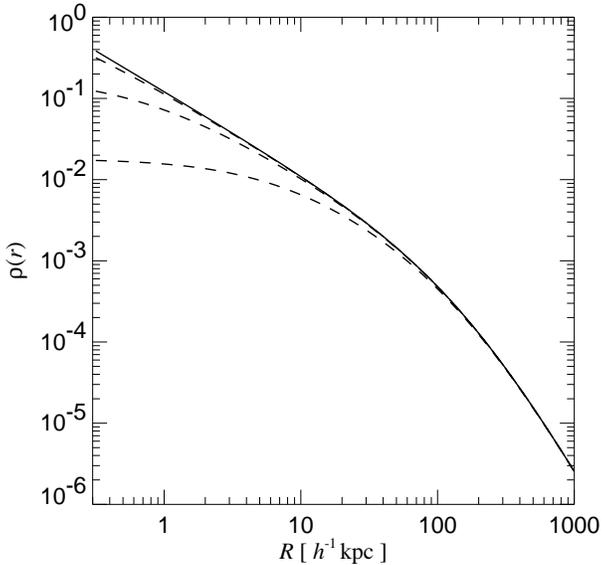}}%
\caption{\label{figprofiles} NFW density profiles used in the test of
time integration criteria. The solid line gives the shape of the
normal (unsoftened) NFW profile, dashed lines are softened according
to $\epsilon= 0.0004$, 0.004, and 0.04, respectively (see equation
\ref{eqnfw}). }  \ec
\end{figure}

We achieve this by `shrinking' the interior of the local
domain. During the neighbour-tree update, we find the maximum spatial
extension of all SPH particles on the local domain. These rectangular
regions are then gathered from all other processors, and cut with the
local extension. If an overlap occurs the local `interior' is reduced
accordingly, thereby resulting in a region that is guaranteed to be
free of any particle lying on another processor.  In this way, the SPH
algorithm will produce correct results, even if the domain
decomposition is repeated only rarely, or never again.  In the case of
periodic boundaries, special care has to be taken in treating cuts of
the current interior with periodic translations of the other domain
extensions.

\section{Results and tests} \label{secres}
	
\subsection{Tests of timestep criteria}

Orbit integration in the collisionless systems of cosmological
simulations is much less challenging than in highly non-linear systems
like star clusters. As a consequence, all the simple timestep criteria
discussed in Section \ref{secticrit} are capable of producing accurate
results for sufficiently small settings of their tolerance
parameters. However, some of these criteria may require a
substantially higher computational effort to achieve a desired level
of integration accuracy than others, and the systematic deviations
under conditions of coarse timestepping may be more severe for certain
criteria than for others.

In order to investigate this issue, we have followed the orbits of the
particle distribution of a typical NFW-halo \citep{NFW,NFW2} for one
Hubble time. The initial radial particle distribution and the initial
velocity distribution were taken from a halo of circular velocity
$1350\,{\rm km\, s^{-1}}$ identified in a cosmological N-body
simulation and well fit by the NFW profile. However, we now evolved
the particle distribution using a fixed softened potential,
corresponding to the mass profile \be \rho(r)= \frac{\delta_c
\rho_{\rm crit.}}{\left(\frac{r}{r_s} +\epsilon\right)
\left(1+\frac{r}{r_s}\right)^2}, \label{eqnfw} \ee fitted to the
initial particle distribution.  We use a static potential in order to
be able to isolate properties of the time integration only. As a
side-effect this also allows a quick integration of all orbits for a
Hubble time, and the analytic formulae for the density and potential
can also be used to check energy conservation for each of the
particles individually.  Note that we introduced a softening parameter
$\epsilon$ in the profile which is not present in the normal NFW
parameterization. The halo had concentration $c=8$ with
$r_s=170\,h^{-1}{\rm kpc}$, and contained about 140000 particles
inside the virial radius.  We followed the orbits of all particles
inside a sphere of radius $3000\lu$ around the halo center (about
200000 in total) using \gadget's integration scheme for single
particles in haloes of three different softening lengths,
$\epsilon=0.0004$, 0.004, and 0.04. The corresponding density profiles
are plotted in Figure~\ref{figprofiles}.

\begin{figure*}
\bc
\resizebox{5.3cm}{!}{\includegraphics{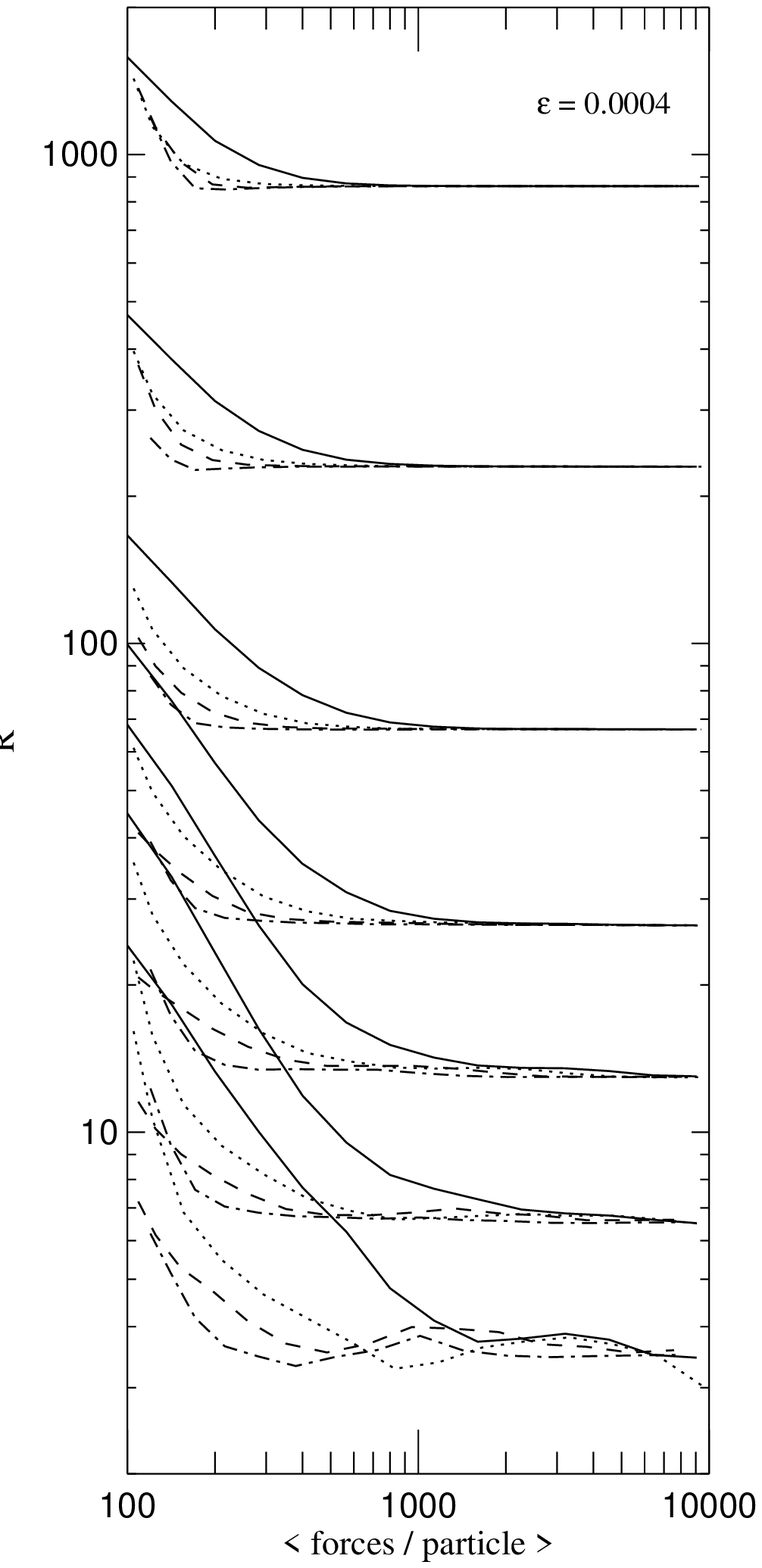}}%
\resizebox{5.3cm}{!}{\includegraphics{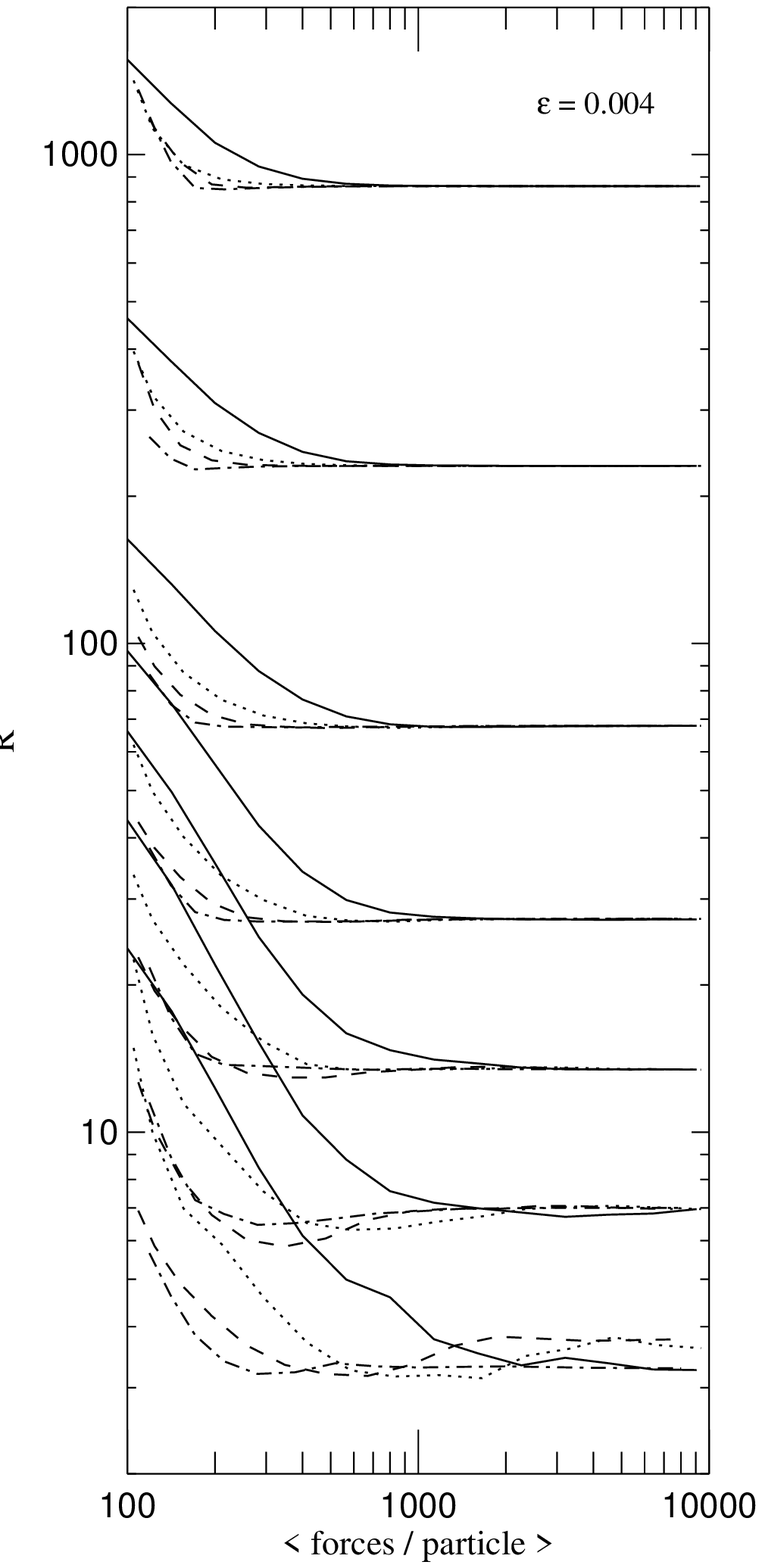}}%
\resizebox{5.3cm}{!}{\includegraphics{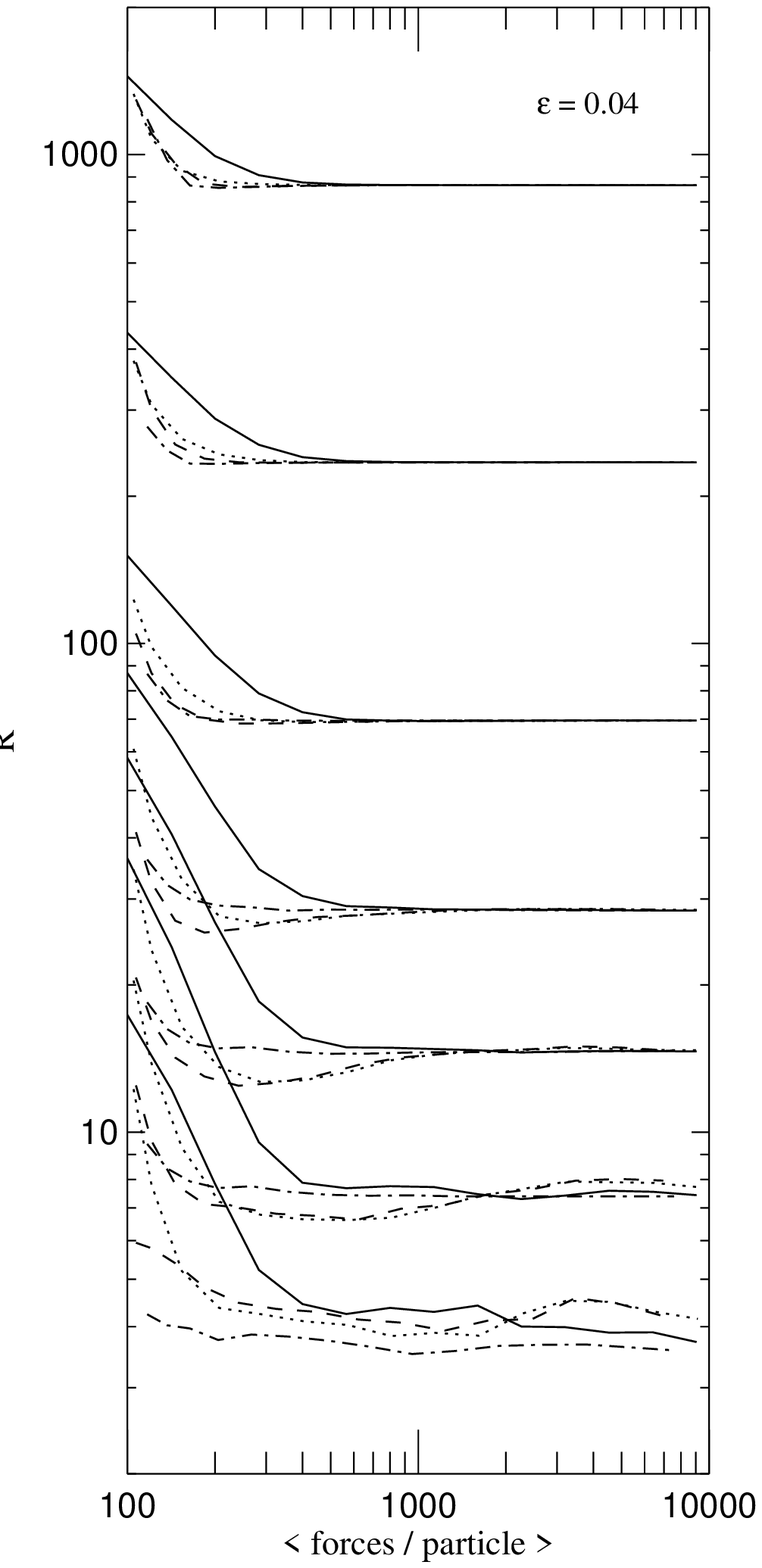}}%
\caption{\label{figshellradii} Radii enclosing a fixed mass
($7.1\times 10^{-5}$, $3.6\times 10^{-4}$, $1.8\times 10^{-3}$,
$7.1\times 10^{-3}$, $3.6\times 10^{-2}$, $2.1\times 10^{-1}$, and
$7.1\times 10^{-1}$ in units of the virial mass) after integrating the
particle population of a NFW halo in a fixed potential for one Hubble
time with various timestep criteria. Solid lines are for fixed
timesteps for all particles, dashed lines for timesteps based on
$\Delta t\propto |\vec{a}|^{-1}$, dotted lines for $\Delta t \propto
|\vec{a}|^{-1/2}$, and dot-dashed for timesteps based on the local
dynamical time, i.e.~$\Delta t\propto \rho^{-1/2}$. The three panels
are for softened forms of the NFW potential according to equation
(\ref{eqnfw}). Note that the fluctuations of the lowest radial shell
are caused by the low sampling of the halo in the core and are not
significant. The horizontal axis gives the mean number of force
evaluation per particle required to advance the particle set for one
Hubble time.} \label{figcrit} \ec
\end{figure*}

In Figure~\ref{figshellradii}, we give the radii of shells enclosing
fixed numbers of particles as a function of the mean number of force
evaluations necessary to carry out the integration for one Hubble
time. The Figure thus shows the convergence of these radii when more
and more computational effort is spent.

The results show several interesting trends. In general, for poor
integration accuracy (on the left), all radii are much larger than the
converged values, i.e.~the density profile shows severe signs of
relaxation even at large distances from the center. When an integrator
with fixed timestep is used (solid lines), the radii decrease
monotonically when more force evaluations, i.e.~smaller timesteps, are
taken, until convergence to asymptotic values is reached. The other
timestep criteria converge to these asymptotic values much more
quickly.  Also, for small softening values, the timestep criterion
$\Delta t\propto |\vec{a}|^{-1}$ (dashed lines) performs noticeably
better than the one with $t\propto |\vec{a}|^{-1/2}$ (dotted lines),
i.e.~it is usually closer to the converged values at a given
computational cost.  On the other hand, the criterion based on the
local dynamical time (dot-dashed lines) does clearly the best job. It
stays close to the converged values even at a very low number of
timesteps.

This impression is corroborated when the softening length is
increased. The timestep based on the local dynamical time performs
better than the other two simple criteria, and the fixed
timestep. Interestingly, the criteria based on the acceleration
develop a non-monotonic behaviour when the potential is softened.
Already at $\epsilon=0.004$ this is visible at the three lowest radii,
but more clearly so for $\epsilon=0.04$. Before the radii increase
when poorer integration settings are chosen, they actually first {\em
decrease}. There is thus a regime where both of these criteria can
lead to a net loss of energy for particles close to the shallow core
induced by the softening, thereby steepening it. We think that this is
related to the fact that the timesteps actually {\em increase} towards
the center for the criteria (\ref{eqtc2}) and (\ref{eqtc3}) as soon as
the logarithmic slope of the density profile becomes shallower than
-1.  On the other hand, the density increases monotonically towards
the center, and hence a timestep based the local dynamical time will
decrease.  If one uses the {\em local} velocity dispersion in the
criterion (\ref{eqtc2}), a better behaviour can be obtained, because
the velocity dispersion declines towards the center in the core region
of dark matter haloes. However, this is not enough to completely
suppress non-monotonic behaviour, as we have found in further sets of
test calculations.

We thus think that in particular for the cores of large haloes the
local dynamical time provides the best of the timestep criteria
tested.  It leads to hardly any secular evolution of the density
profile down to rather coarse timestepping. On the other hand, as
\citet{Qu97} have pointed out, estimating the density in a real
simulation poses additional problems. Since kernel estimates are
carried out over a number of neighbours, the density in small haloes
can be underestimated, resulting in `evaporation' of haloes. These
haloes are better treated with a criterion based on local
acceleration, which is much more accurately known in this case.

We thus think that a conservative and accurate way to choose timesteps
in cosmological simulations is obtained by taking the minimum of
(\ref{eqtc2}) and (\ref{eqtc4}). This requires a computation of the
local matter density and local velocity dispersion of collisionless
material.  Both of these quantities can be obtained for the dark
matter just as it is done for the SPH particles. Of course, this
requires additional work in order to find neighbours and to do the
appropriate summations and kernel estimates, which is, however,
typically not more than $\sim 30$\% of the cost the gravitational
force computation.  However, as Figure~\ref{figcrit} shows, this is in
general worth investing, because the alternative timestep criteria
require more timesteps, and hence larger computational cost, by a
larger factor in order to achieve the same accuracy.

\begin{figure*}
\bc 
\parbox{16.1cm}{\resizebox{8.0cm}{!}{\includegraphics{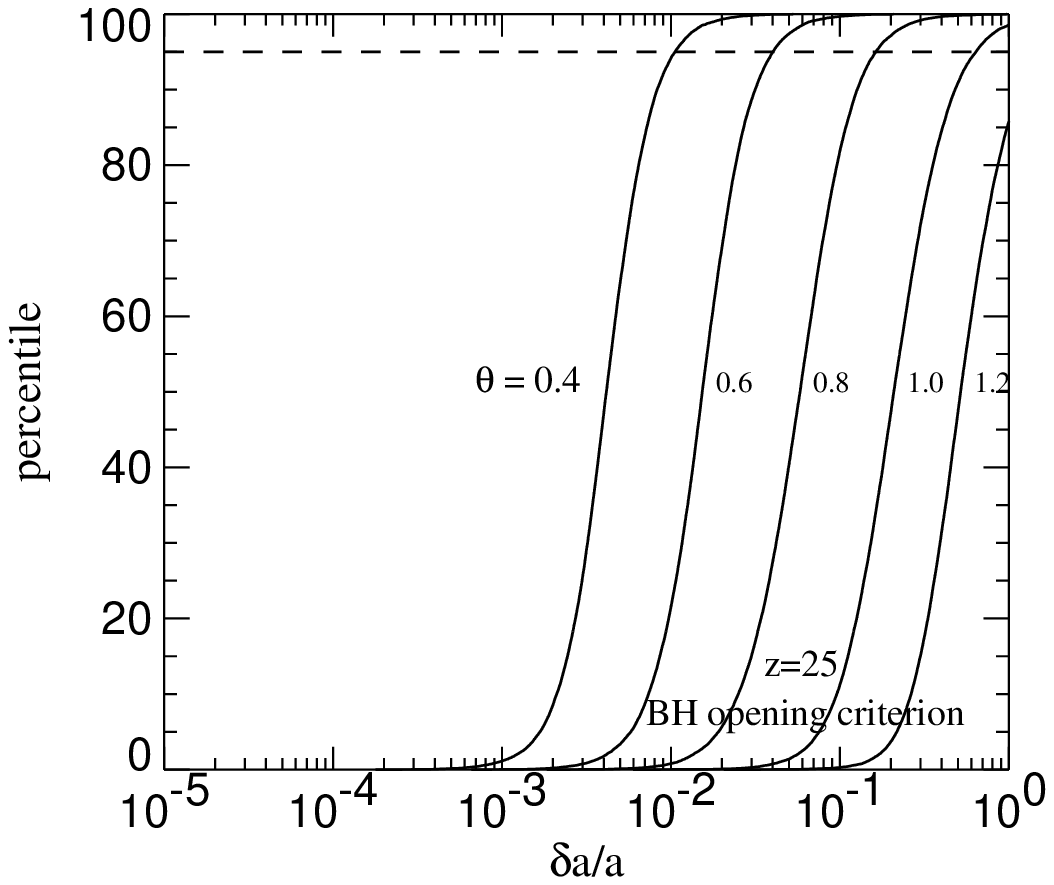}}%
\resizebox{8.0cm}{!}{\includegraphics{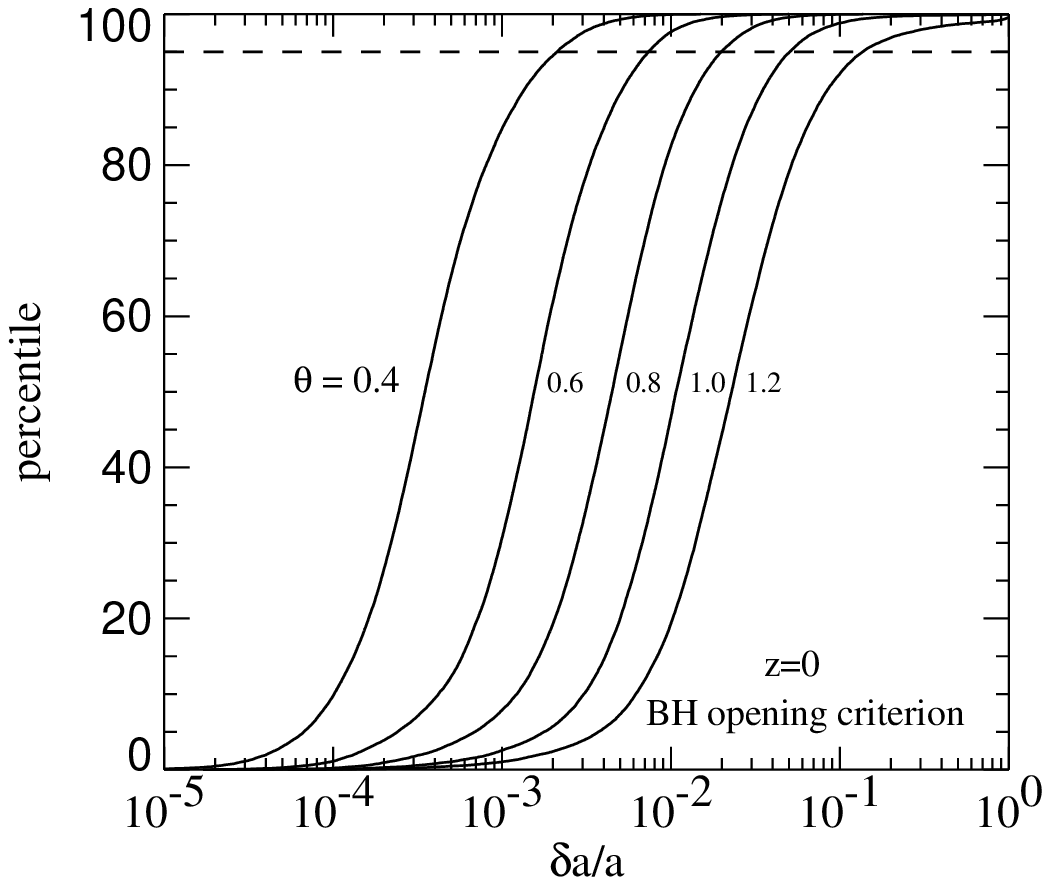}}\\%
\resizebox{8.0cm}{!}{\includegraphics{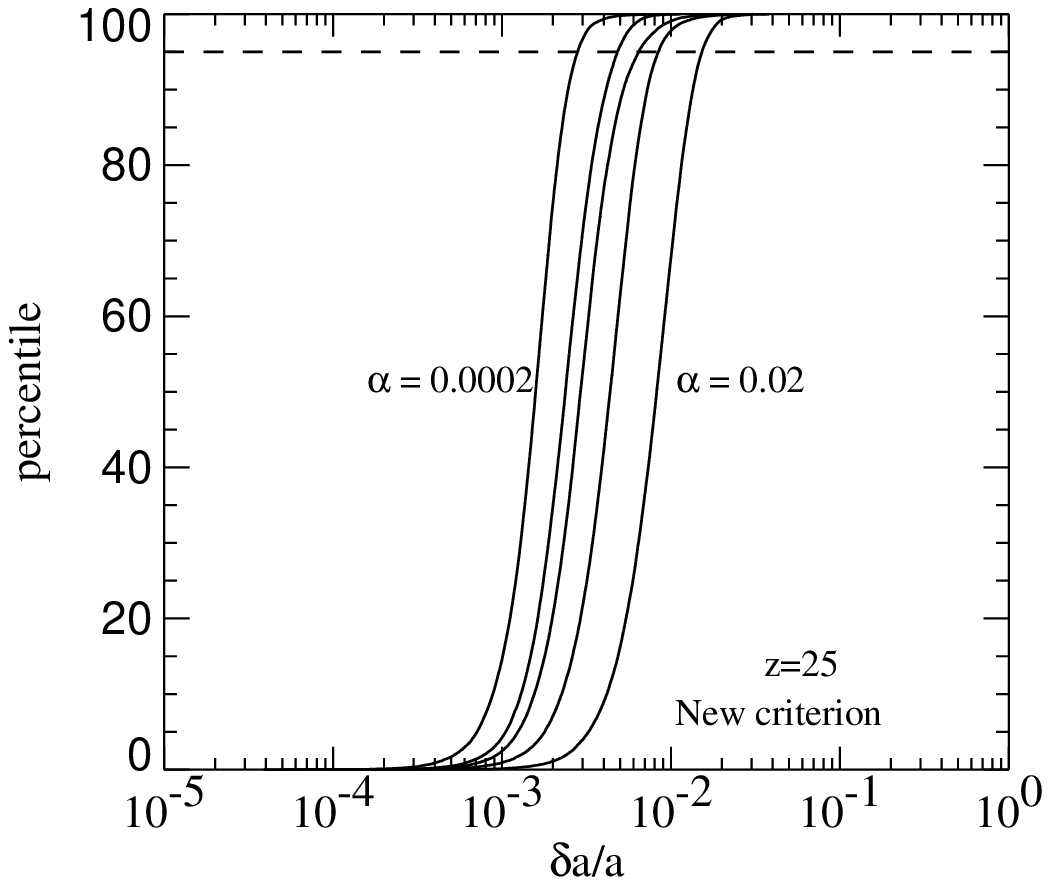}}%
\resizebox{8.0cm}{!}{\includegraphics{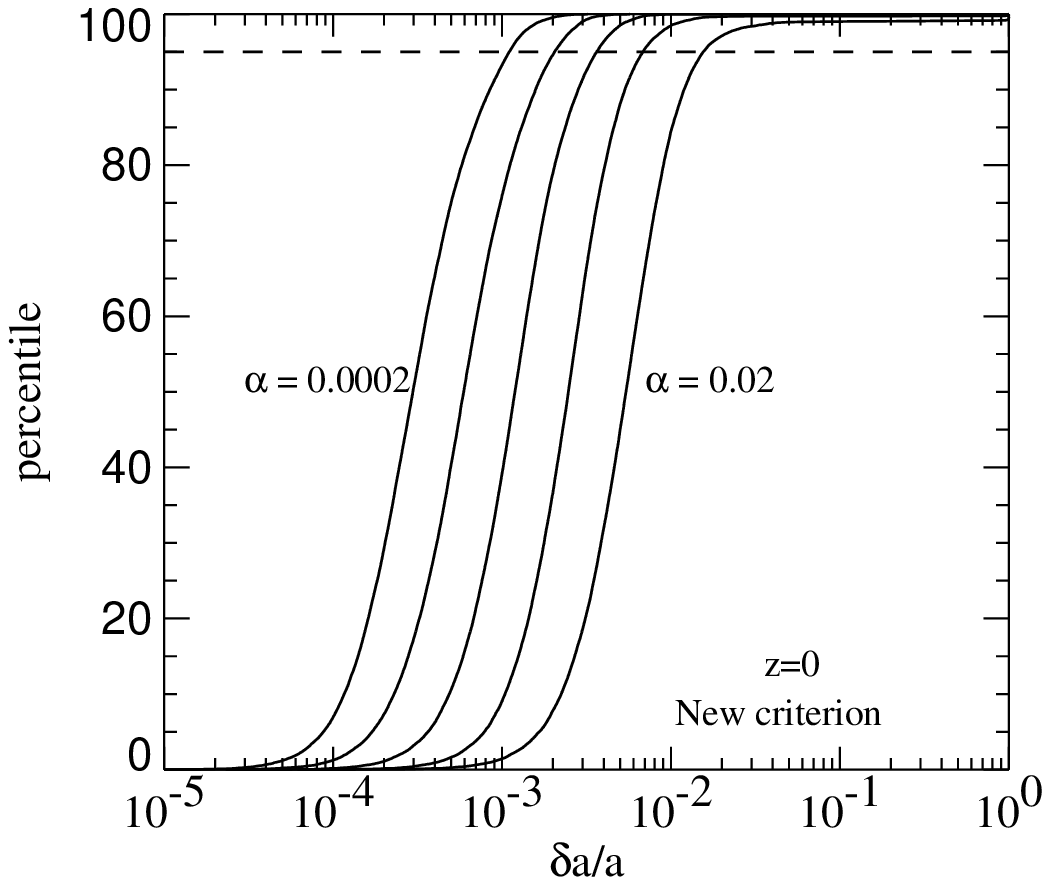}}\\%
\resizebox{8.0cm}{!}{\includegraphics{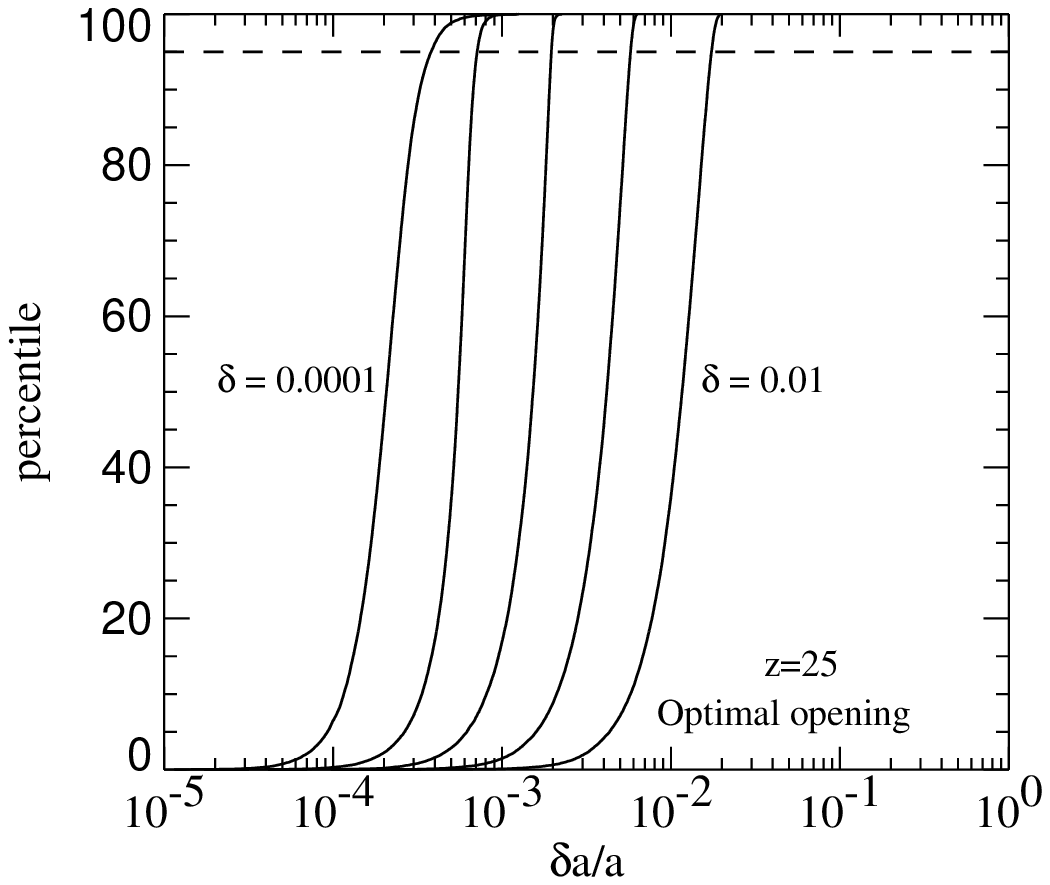}}%
\resizebox{8.0cm}{!}{\includegraphics{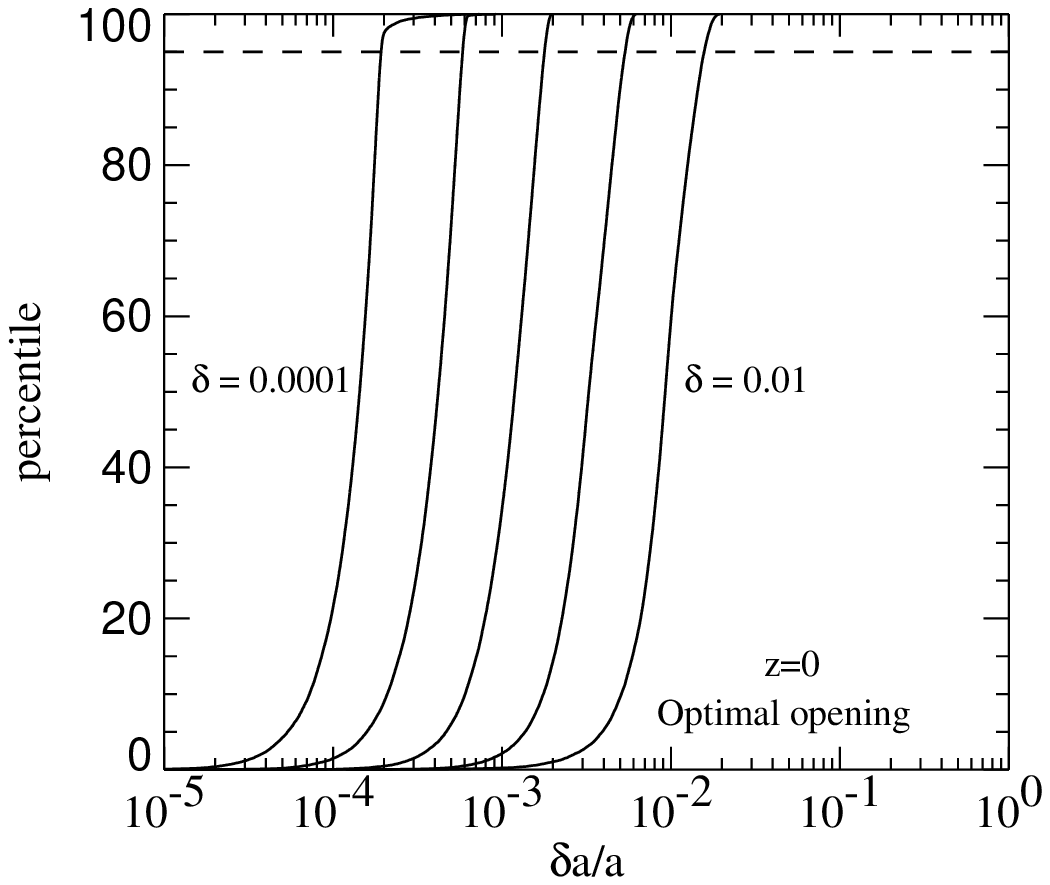}}\\%
}
\caption
{Cumulative distribution of the relative force error obtained as a
function of the tolerance parameter for three opening criteria, and for
two different particle distribution. The panels in the left column
show results for the particle distribution in the initial conditions
of a $32^3$ simulation at $z=25$, while the panels on the right give
the force errors for the evolved and clustered distribution at $z=0$.
\label{cumulerrors}}
\vspace*{0.5cm}\ \\
\ec
\end{figure*}

\begin{figure*}
\bc
\parbox{16.1cm}{
\resizebox{8.1cm}{!}{\includegraphics{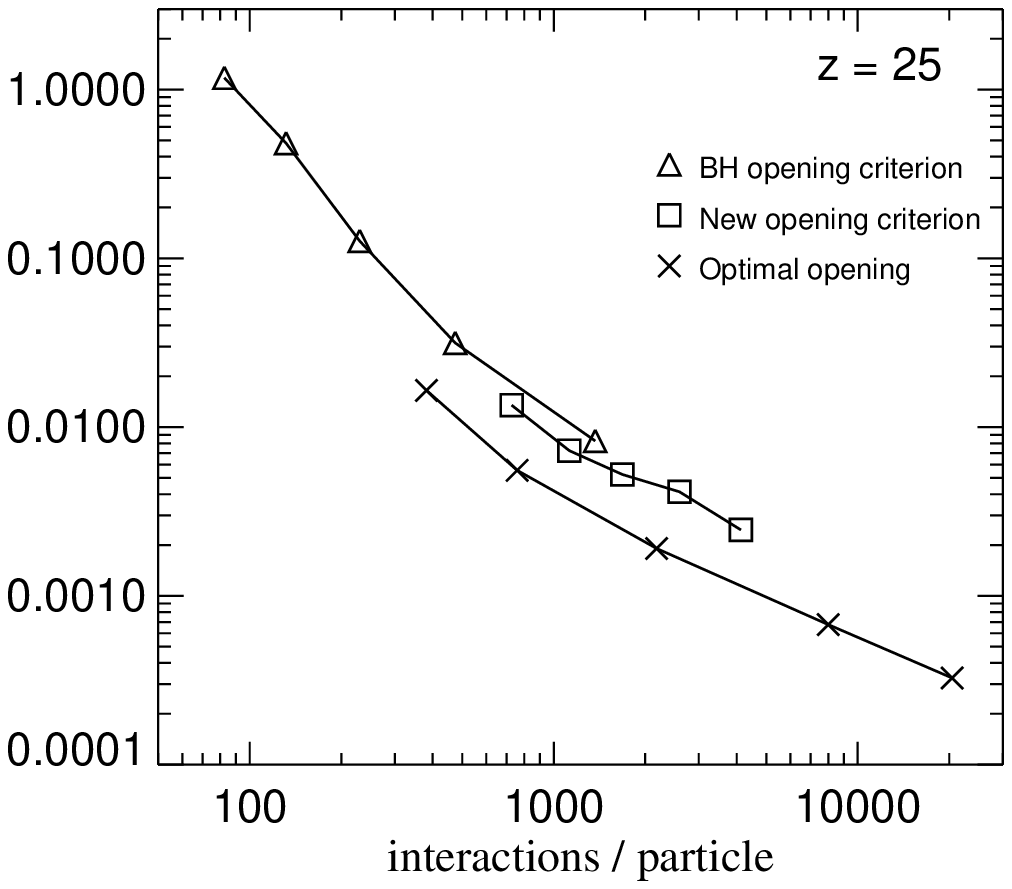}}%
\resizebox{8.1cm}{!}{\includegraphics{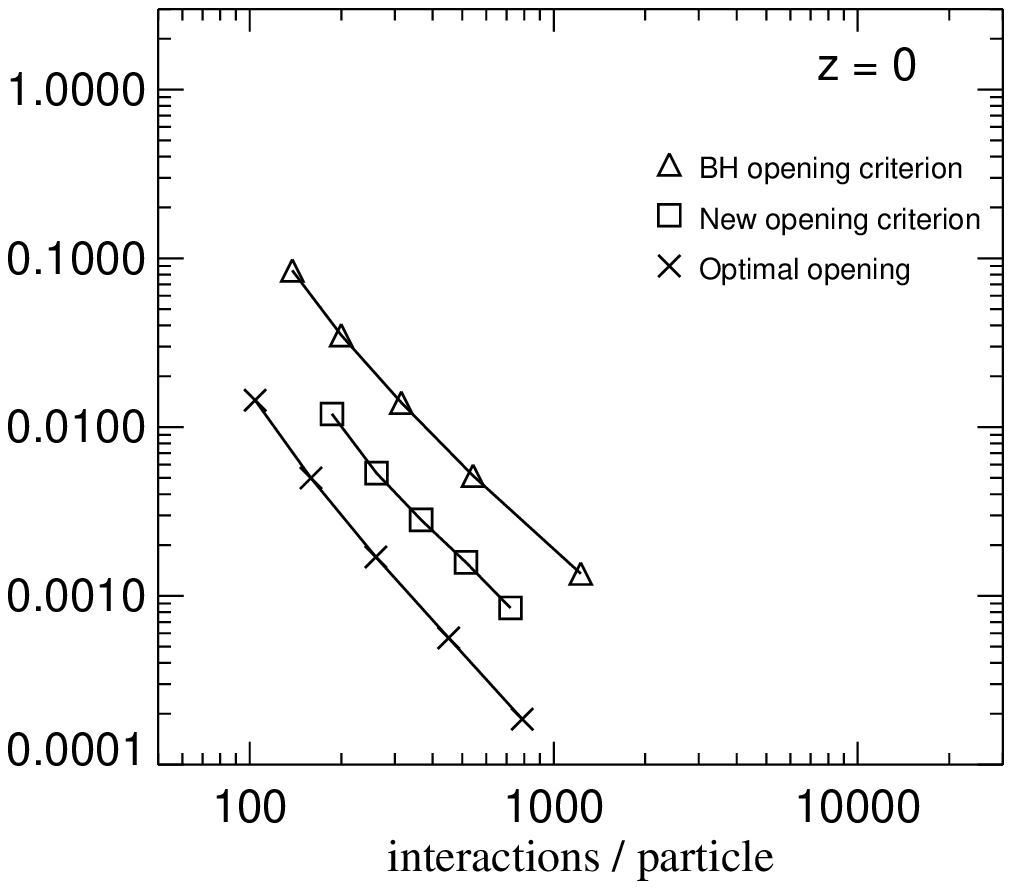}}\\%
}
\caption
{Computational efficiency of three different cell-opening criteria,
parameterized by their tolerance parameters. The horizontal axes is
proportional to the computational cost, and the vertical axes gives
the 95\% percentile of the cumulative distribution of relative force
errors in a $32^3$ cosmological simulation. The left panel shows
results for the initial conditions at $z=25$, the right panel for the
clustered distribution at redshift $z=0$.
\label{performplot}}
\ec
\end{figure*}

\subsection{Force accuracy and opening criterion\label{newopsecresults}}

In Section~\ref{newopsec}, we have outlined that standard values of
the BH-opening criterion can result in very high force errors for the
initial conditions of cosmological N-body simulations. Here, the
density field is very close to being homogeneous, so that small
peculiar accelerations arise from a cancellation of relatively large
partial forces.  We now investigate this problem further using a
cosmological simulation with $32^3$ particles in a periodic box.  We
consider the particle distribution of the initial conditions at
redshift $z=25$, and the clustered one of the evolved simulation at
$z=0$. Romeel Dav\'{e} has kindly made these particle dumps available
to us, together with exact forces he calculated for all particles with
a direct summation technique, properly including the periodic images
of the particles.  In the following we will compare his forces (which
we take to be the exact result) to the ones computed by the parallel
version of \gadget\ using two processors on a Linux PC.

We first examine the distribution of the relative force error as a
function of the tolerance parameter used in the two different
cell-opening criteria (\ref{eqopen}) and (\ref{eqnewcrit}). These
results are shown in Figure \ref{cumulerrors}, where we also contrast
these distributions with the ones obtained with an {\em optimum}
cell-opening strategy. The latter may be operationally defined as
follows:

Any complete tree walk results in an interaction list that contains a
number of internal tree nodes (whose multipole expansions are used)
and a number of single particles (which give rise to exact partial
forces). Obviously, the shortest possible interaction list is that of
just one entry, the root node of the tree itself. Suppose we start
with this interaction list, and then open always the one node of the
{\em current} interaction list that gives rise to the largest absolute
force error. This is the node with the largest difference between
multipole expansion and exact force, as resulting from direct
summation over all the particles represented by the node. With such an
opening strategy, the shortest possible interaction list can be found
for any desired level of accuracy. The latter may be set by requiring
that the largest force error of any node in the interaction list has
fallen below a fraction $\delta |\vec{a}|$ of the true force field.

Of course, this optimum strategy is not practical in a real
simulation, after all it requires the knowledge of all the exact
forces for all internal nodes of the tree. If these were known, we
wouldn't have to bother with trees to begin with.  However, it is very
interesting to find out what such a fiducial optimum opening strategy
would ultimately deliver.  Any other analytic or heuristic
cell-opening criterion is bound to be worse. Knowing how much worse
such criteria are will thus inform us about the maximum performance
improvement possible by adopting alternative cell-opening criteria.
We therefore computed for each particle the exact direct summation
forces exerted by each of the internal nodes, hence allowing us to
perform an optimum tree walk in the above way. The resulting
cumulative error distributions as a function of $\delta$ are shown in
Figure~\ref{cumulerrors}.

Looking at this Figure, it is clear that the BH error distribution has
a more pronounced tail of large force errors compared to the opening
criterion (\ref{eqnewcrit}). What is most striking however is that at
high redshift the BH force errors can become very large for values of
the opening angle $\theta$ that tend to give perfectly acceptable
accuracy for a clustered mass distribution. This is clearly caused by
its purely geometrical nature, which does not know about the smallness
of the net forces, and thus does not invest sufficient effort to evaluate
partial forces to high enough accuracy.

The error distributions alone do not tell yet about the computational
efficiency of the cell-opening strategies. To assess these, we define
the error for a given cell-opening criterion as the 95\% percentile of
the error distribution, and we plot this versus the mean length of the
interaction list per particle. This length is essentially linearly
related to the computational cost of the force evaluation.

In Figure~\ref{performplot}, we compare the performance of the
cell-opening criteria in this way. At high redshift, we see that the
large errors of the BH criterion are mainly caused because the mean
length of the interaction list remains small and does not adapt to the
more demanding requirements of the force computation in this
regime. Our `new' cell-opening criterion does much better in this
respect; its errors remain well controlled without having to adjust
the tolerance parameter.

\begin{figure}
\bc
\resizebox{8cm}{!}{\includegraphics{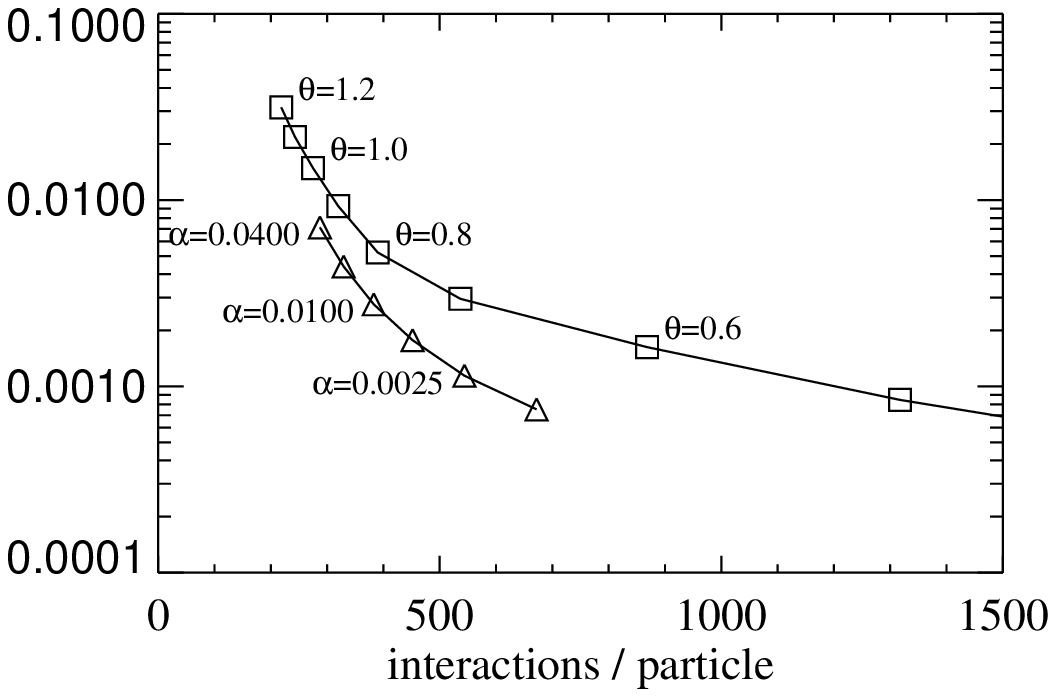}}
\caption
{Force errors for an isolated halo/disk galaxy with the BH-criterion
(boxes), and the new opening criterion (triangles). The dark halo of
the galaxy is modeled with a NFW profile and is truncated at the
virial radius. The plot shows the 90\% percentile of the error
distribution, i.e.\ 90\% of the particles have force errors below the
cited values. The horizontal axes measures the computational expense.
\label{figF2}}
\ec
\end{figure}

Another advantage of the new criterion is that it is in fact more
efficient than the BH criterion, i.e.~it achieves higher force
accuracy for a given computational expense. As Figure
\ref{performplot} shows, for a clustered particle distribution in a
cosmological simulation, the implied saving can easily reach a factor
2-3, speeding up the simulation by the same factor. Similar
performance improvements are obtained for isolated galaxies, as the
example in Figure~\ref{figF2} demonstrates.  This can be understood as
follows: The geometrical BH criterion does not take the dynamical
significance of the mass distribution into account. For example, it
will invest a large number of cell-particle interactions to compute
the force exerted by a large void to a high relative accuracy, while
actually this force is of small absolute size, and it would be better
to concentrate more on those regions that provide most of the force on
the current particle.  The new opening criterion follows this latter
strategy, improving the force accuracy at a given number of
particle-cell interactions.

We note that the improvement obtained by criterion (\ref{eqnewcrit})
brings us about halfway to what might ultimately be possible with an
optimum criterion.  It thus appears that there is still room for
further improvement of the cell-opening criterion, even though it is
clear that the optimum will likely not be reachable in practice.

\subsection{Colliding disk galaxies}

As a test of the performance and accuracy of the integration of
collisionless systems, we here consider a pair of merging disk
galaxies.  Each galaxy has a massive dark halo consisting of 30000
particles, and an embedded stellar disk, represented by 20000
particles. The dark halo is modeled according to the NFW-profile,
adiabatically modified by the central exponential disk, which
contributes 5\% of the total mass.  The halo has a circular velocity
$v_{200}=160\,{\rm km}\,{\rm s}^{-1}$, a concentration $c=5$, and spin
parameter $\lambda=0.05$. The radial exponential scale length of the
disk is $R_{\rm d}=4.5\lu$, while the vertical structure is that of an
isothermal sheet with thickness $z_0=0.2R_{\rm d}$. The gravitational
softening of the halo particles is set to $0.4\lu$, and that of the
disk to $0.1\lu$.

Initially, the two galaxies are set-up on a parabolic orbit, with
separation such that their dark haloes just touch. Both of the
galaxies have a prograde orientation, but are inclined with respect to
the orbital plane.  In fact, the test considered here corresponds
exactly to the simulation `C1' computed by \citet{Spr99}, where more
information about the construction of the initial conditions can be
found \citep[see also][]{Sp98}.

We first consider a run of this model with a set of parameters equal
to the coarsest values we would typically employ for a simulation of
this kind.  For the time integration, we used the parameter
$\alpha_{\rm tol}\sigma= 25\,{\rm km}\,{\rm s}^{-1}$, and for the
force computation with the tree algorithm, the new opening criterion
with $\alpha=0.04$. The simulation was then run from $t=0$ to $t=2.8$
in internal units of time (corresponding to $2.85\,h^{-1}{\rm
Gyr}$). During this time the galaxies have their first close encounter
at around $t\simeq 1.0$, where tidal tails are ejected out of the
stellar disks. Due to the braking by dynamical friction, the galaxies
eventually fall together for a second time, after which they quickly
merge and violently relax to form a single merger remnant. At $t=2.8$,
the inner parts of the merger remnant are well relaxed.

This simulation required a total number of 4684 steps and 27795733
force computations, i.e.~a computationally equally expensive
computation with fixed timesteps could just make 280 full timesteps.
The relative error in the total energy was $3.0\times 10^{-3}$, and a
Sun Ultrasparc-II workstation (sparcv9 processor, 296 MHz clock speed)
did the simulation in 4 hours (using the serial code). The raw force
speed was $\sim$2800 force computations per second, older workstations
will achieve somewhat smaller values, of course. Also, higher force
accuracy settings will slow down the code somewhat.

\begin{figure}
\bc
\resizebox{8cm}{!}{\includegraphics{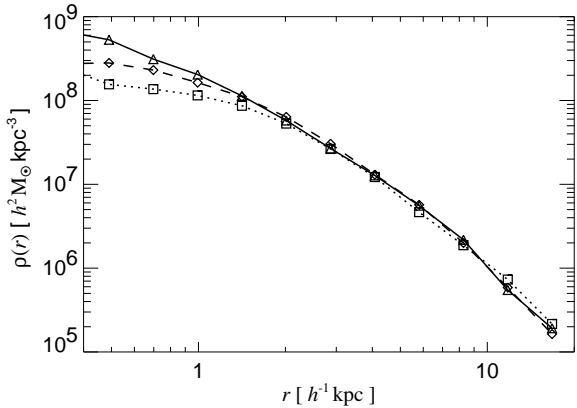}}
\caption
{Spherically averaged density profile of the stellar component in the
merger remnant of two colliding disk galaxies.  Triangles show the
results obtained using our variable timestep integration, while boxes
and diamonds are for fixed timestep integrations with $\Delta t= 0.01$
($10\,h^{-1}{\rm Myr}$) and $\Delta t= 0.0025$ ($2.5\,h^{-1}{\rm
Myr}$), respectively.  Note that the simulation using the adaptive
timestep is about as expensive as the one with $\Delta t= 0.01$.  In
each case, the center of the remnant was defined as the position of
the particle with the minimum gravitational potential.
\label{densprofcomp}}
\ec
\end{figure}

We now consider a simulation of the same system using a fixed
timestep.  For $\Delta t=0.01$, the run needs 280 full steps, i.e. it
consumes the same amount of CPU time as above. However, in this
simulation, the error in the total energy is $2.2\%$, substantially
larger than before.  There are also differences in the density profile
of the merger remnants.  In Figure~\ref{densprofcomp}, we compare the
inner density profile of the simulation with adaptive timesteps
(triangles) to the one with a fixed timestep of $\Delta t=0.01$
(boxes). We here show the spherically averaged profile of the stellar
component, with the center of the remnants defined as the position of
the particle with the minimum gravitational potential.  It can be seen
that in the innermost $\sim 1 \lu$, the density obtained with the
fixed timestep falls short of the adaptive timestep integration.

To get an idea how small the fixed timestep has to be to achieve
similar accuracy as with the adaptive timestep, we have simulated this
test a second time, with a fixed timesteps of $\Delta t=0.0025$. We
also show the corresponding profile (diamonds) in
Figure~\ref{densprofcomp}. It can be seen that for smaller $\Delta t$,
the agreement with the variable timestep result improves.  At $\sim
2\times 0.4\lu$, the softening of the dark matter starts to become
important.  For agreement down to this scale, the fixed timestep needs
to be slightly smaller than $\Delta t = 0.0025$, so the overall saving
due to the use of individual particle timesteps is at least a factor
of $4-5$ in this example.

\begin{figure*}
\bc
\resizebox{14.0cm}{!}{\includegraphics{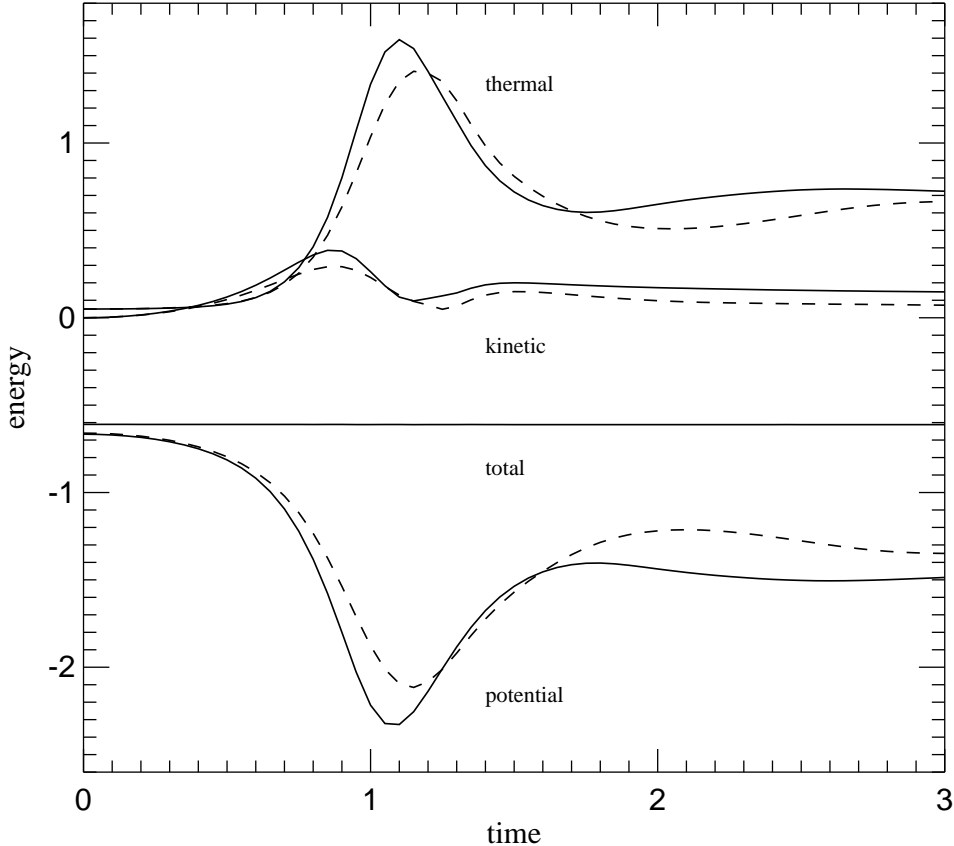}}
\caption
{Time evolution of the thermal, kinetic, potential, and total energy
for the collapse of an initially isothermal gas sphere.  Solid lines
show results for a simulations with 30976 particles, dashed lines are
for a 4224 particle run.
\label{figgassphere}}
\ec
\end{figure*}

\subsection{Collapse of a cold gas sphere}

The self-gravitating collapse of an initially isothermal, cool gas
sphere has been a common test problem of SPH codes \citep[and
others]{Ev88,He89,St93,Th98,Ca98}. Following these authors, we
consider a spherically symmetric gas cloud of total mass $M$, radius
$R$, and initial density profile
\be
\rho(r)=\frac{M}{2\pi R^2}\;\frac{1}{r}.
\label{eqp1}
\ee 
We take the gas to be of constant temperature initially, with an
internal energy per unit mass of
\be
u=0.05\,\frac{GM}{R}.
\ee
At the start of the simulation, the gas particles are at rest. We
obtain their initial coordinates from a distorted regular grid that
reproduces the density profile (\ref{eqp1}), and we use a system of
units with $G=M=R=1$.

In Figure \ref{figgassphere}, we show the evolution of the potential,
the thermal, and the kinetic energy of the system from the start of
the simulation at $t=0$ to $t=3$. We plot results for two simulations,
one with 30976 particles (solid), and one with 4224 particles
(dashed).  During the central bounce between $t\approx 0.8$ and
$t\approx 1.2$ most of the kinetic energy is converted into heat, and
a strong shock wave travels outward.  Later, the system slowly settles
to virial equilibrium.

For these runs $N_{\rm s}$ was set to 40, the gravitational softening
to $\epsilon=0.02$, and time integration was controlled by the
parameters $\alpha_{\rm tol}\sigma=0.05$ and $\alpha_{\rm
cour}=0.1$\footnote{Note that our definition of the smoothing length
$h$ differs by a factor of 2 from most previous SPH
implementations. As a consequence, corresponding values of
$\alpha_{\rm cour}$ are different by a factor of 2, too.}, resulting
in very good energy conservation. The absolute error in the total
energy is less than $1.1\times 10^{-3}$ at all times during the
simulation, translating to a relative error of 0.23\%. Since we use a
time integration scheme with individual timesteps of arbitrary size,
this small error is reassuring.  The total number of small steps taken
by the 4224 particle simulation was 3855, with a total of 2192421
force computations, i.e. the equivalent number of `full' timesteps was
519. A Sun Ultrasparc-II workstation needed 2300 seconds for the
simulation.  The larger 30976 particle run took 10668 steps, with an
equivalent of 1086 `full' steps, and 12 hours of CPU time. Note that
by reducing the time integration accuracy by a factor of 2, with a
corresponding reduction of the CPU time consumption by the same
factor, the results do practically not change, however, the maximum
error in the energy goes up to 1.2\% in this case.

\begin{table*}
\caption{\label{tab1}Consumption of CPU-time in various parts of the code
for a cosmological run from $z=50$ to $z=4.3$.  The table gives timings for
runs with 4, 8 and 16 processors on the Garching Cray T3E. The
computation of the gravitational forces is by far the dominant
computational task. We have further split up that time into the actual
tree-walk time, the tree-construction time, the time for communication
and summation of force contributions, and into the time lost by
work-load imbalance.  The potential computation is done only once in
this test (it can optionally be done in regular intervals to check
energy conservation of the code). `Miscellaneous' refers to time spent
in advancing and predicting particles, and in managing the binary tree
for the timeline. I/O time for writing a snapshot file (groups of
processors can write in parallel) is only 1-2 seconds, and therefore
not listed here.}
\bc
\begin{tabular}{lrrrrrr}
\hline
 & \multicolumn{2}{c}{\ \hspace{0.9cm}{\bf 4 processors}} &  \multicolumn{2}{c}{\ \hspace{0.9cm}{\bf 8
 processors}} & \multicolumn{2}{c}{\ \hspace{0.9cm}{\bf 16 processors}} \\
  & \parbox[t]{2.0cm}{\flushright cumulative\\time [sec]} &
  \parbox[t]{1.3cm}{\flushright relative\\time} & 
 \parbox[t]{2.0cm}{\flushright cumulative\\time [sec]} &
 \parbox[t]{1.3cm}{\flushright relative\\time} & 
 \parbox[t]{2.0cm}{\flushright cumulative\\time [sec]} &
 \parbox[t]{1.3cm}{\flushright relative\\time} \\
\hline
total 	& 8269.0 & 100.0 \% & 4074.0 & 100.0 \% & 2887.5 & 100.0 \% \\
gravity & 7574.6 & 91.6 \% & 3643.0& 89.4 \% & 2530.3 & 87.6 \% \\
{\small\hspace{0.5cm}tree walks} & 5086.3 & 61.5  \% &  2258.4 & 55.4
\% &  1322.4 &  45.8  \% \\
{\small\hspace{0.5cm}tree construction} & 1518.4&  18.4  \% &  773.7 & 19.0\% &  588.4 &  20.4  \% \\
{\small\hspace{0.5cm}communication \& summation } & 24.4 & 0.3  \% & 35.4  & 0.9  \%  & 54.1  & 1.9  \%\\
{\small\hspace{0.5cm}work-load imbalance}  & 901.5 &  10.9 \%  & 535.1  &  13.1 \%  &   537.4 & 18.6 \% \\
domain decomposition & 209.9  & 2.5 \%   & 158.1  & 3.9 \% & 172.2  & 6.0 \% \\
potential  computation (optional) & 46.3   & 0.2 \% & 18.1   & 0.4 \% & 11.4   & 0.4 \% \\
miscellaneous  & 438.3  & 5.3 \% & 254.8  & 6.3 \% & 173.6  & 6.0 \% \\
\hline
\end{tabular}
\ec
\end{table*}

The results of Figure~\ref{figgassphere} agree very well with those of
\citet{St93}, and with \cite{Th98} if we compare to their `best'
implementation of artificial viscosity (their version 12).
\citet{St93} have also computed a solution of this problem with a very
accurate, one-dimensional, spherically symmetric piecewise parabolic
method (PPM). For particle numbers above 10000, our SPH results become
very close to the finite difference result. However, even for very
small particle numbers, SPH is capable of reproducing the general
features of the solution very well. We also expect that a {\em
three-dimensional} PPM calculation of the collapse would likely
require at least the same amount of CPU time as our SPH calculation.

\subsection{Performance and scalability of the parallel gravity}

We here show a simple test of the performance of the parallel version
of the code under conditions relevant for real target applications.
For this test, we have used a `stripped-down' version of the initial
conditions originally constructed for a high-resolution simulation of
a cluster of galaxies.  The original set of initial conditions was
set-up to follow the cosmological evolution of a large spherical
region with comoving radius $70\Mlu$, within a $\Lambda$CDM cosmogony
corresponding to $\Omega_0=0.3$, $\Omega_\Lambda=0.7$, $z_{\rm
start}=50$, and $h=0.7$.  In the center of the simulation sphere, 2
million high-resolution particles were placed in the somewhat enlarged
Lagrangian region of the cluster. The rest of the volume was filled
with an extended shell of boundary particles of larger mass and larger
softening; they are needed for a proper representation of the
gravitational tidal field.

\begin{figure*}
\bc
\resizebox{8cm}{!}{\includegraphics{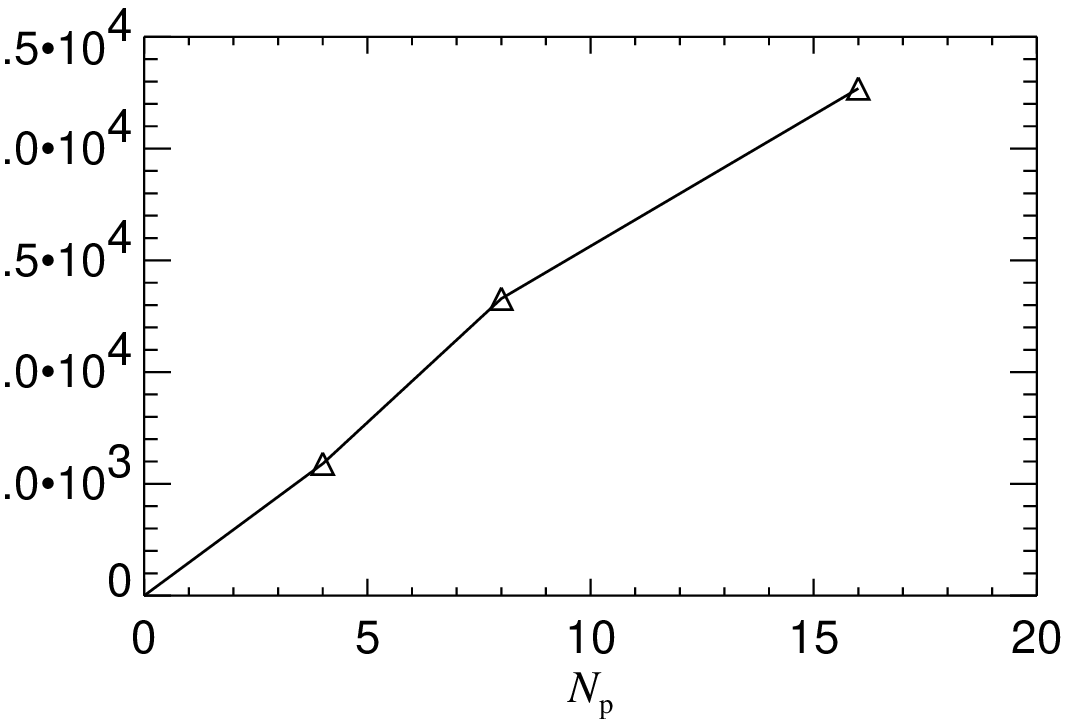}}
\resizebox{8cm}{!}{\includegraphics{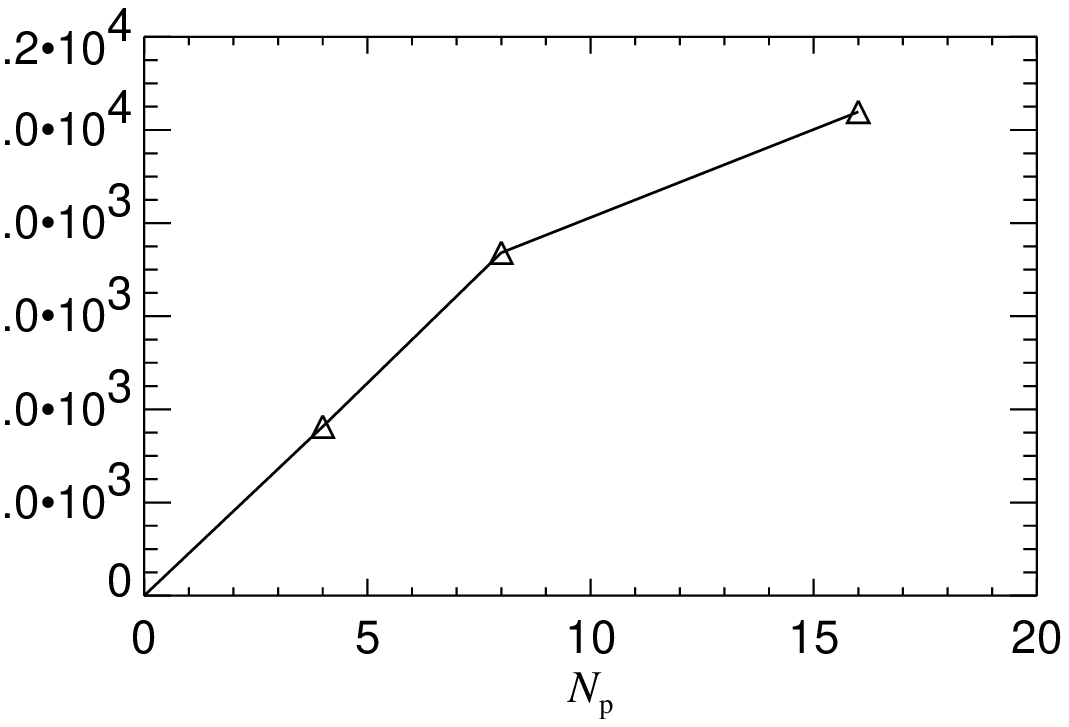}}
\caption{\label{fig1} Code performance and scalability for a
cosmological integration with vacuum boundaries. The left panel shows
the speed of the gravitational force computation as a function of
processor number (in particles per second).  This is based on the tree
walk time alone. In the simulation, additional time is needed for tree
construction, work-load imbalance, communication, domain
decomposition, prediction of particles, timeline, etc.  This reduces
the `effective' speed of the code, as shown in the right panel.  This
effective speed gives the number of particles advanced by one timestep
per second.  Note that the only significant source of work-load
imbalance in the code occurs in the gravity computation, where some
small fraction of time is lost when processors idly wait for others to
finish their tree-walks.}  
\ec
\end{figure*}

To keep our test simple, we have cut a centred sphere of comoving
radius $12.25 \Mlu$ from these initial conditions, and we only
simulated the 500000 high-resolution particles with mass $m_{\rm
p}=1.36\times 10^{9}h^{-1}\msun$ found within this region. Such a
simulation will not be useful for direct scientific analysis because
it does not model the tidal field properly. However, this test will
show realistic clustering and time-stepping behaviour, and thus allows
a reasonable assessment of the expected computational cost and scaling
of the full problem.

We have run the test problem with \gadget\ on the Garching T3E from
redshift $z=50$ to redshift $z=4.3$. We repeated the identical run on
partitions of size 4, 8, and 16 processors.  In this test, we included
quadrupole moments in the tree computation, we used a BH opening
criterion with $\theta=1.0$, and a gravitational softening length of
$15\lu$.

In Table \ref{tab1} we list in detail the elapsed wall-clock time for
various parts of the code for the three simulations.  The dominant
sink of CPU time is the computation of gravitational forces for the
particles. To advance the test simulation from $z=50$ to $z=4.3$,
\gadget\ needed $30.0\times 10^6$ force computations in a total of
3350 timesteps. Note that on average only 1.8\% of the particles are
advanced in a single timestep.  Under these conditions it is
challenging to eliminate sources of overhead incurred by the
time-stepping and to maintain work-load balancing. \gadget\ solves
this task satisfactorily.  If we used a fixed timestep, the work-load
balancing would be essentially perfect. Note that the use of our
adaptive timestep integrator results in a saving of about a factor of
3-5 compared to a fixed timestep scheme with the same accuracy.

We think that the overall performance of \gadget\ is good in this
test.  The raw gravitational speed is high, and the algorithm used to
parallelize the force computation scales well, as is seen in the left
panel of Figure \ref{fig1}. Note that the force-speed of the $N_{\rm
p}=8$ run is even higher than that of the $N_{\rm p}=4$ run. This is
because the domain decomposition does exactly one split in the $x$-,
$y$-, and $z$-directions in the $N_{\rm p}=8$ case.  The domains are
then close to cubes, which reduces the depth of the tree and speeds up
the tree-walks.

Also, the force communication does not involve a significant
communication overhead, and the time spent in miscellaneous tasks of
the simulation code scales closely with processor number.  Most losses
in \gadget\ occur due to work-load imbalance in the force computation.
While we think these losses are acceptable in the above test, one
should keep in mind that we here kept the problem size {\em fixed},
and just increased the processor number.  If we {\em also scale up the
problem size}, work-load balancing will be significantly easier to
achieve, and the efficiency of \gadget\ will be nearly as high as for
small processor number.

Nevertheless, the computational speed of \gadget\ may seem
disappointing when compared with the theoretical peak performance of
modern microprocessors. For example, the processors of the T3E used
for the timings have a nominal peak performance of 600 MFlops, but
\gadget\ falls far short of reaching this. However, the peak
performance can only be reached under the most favourable of
circumstances, and typical codes operate in a range where they are a
factor of 5-10 slower. \gadget\ is no exception here. While we think
that the code does a reasonable job in avoiding unnecessary floating
point operations, there is certainly room for further tuning measures,
including processor-specific ones which haven't been tried at all so
far. Also note that our algorithm of individual tree walks produces
essentially random access of memory locations, a situation that could
hardly be worse for the cache pipeline of current microprocessors.

\subsection{Parallel SPH in a periodic volume}

As a further test of the scaling and performance of the parallel
version of \gadget\ in typical cosmological applications we consider a
simulation of structure formation in a periodic box of size
$(50h^{-1}{\rm Mpc})^3$, including adiabatic gas physics.  We use
$32^3$ dark matter particles, and $32^3$ SPH particles in a
$\Lambda$CDM cosmology with $\Omega_0=0.3$, $\Omega_\Lambda=0.7$ and
$h=0.7$, normalized to $\sigma_8=0.9$. For simplicity, initial
conditions are constructed by displacing the particles from a grid,
with the SPH particles placed on top of the dark matter particles.

\begin{figure}
\bc
\resizebox{8.0cm}{!}{\includegraphics{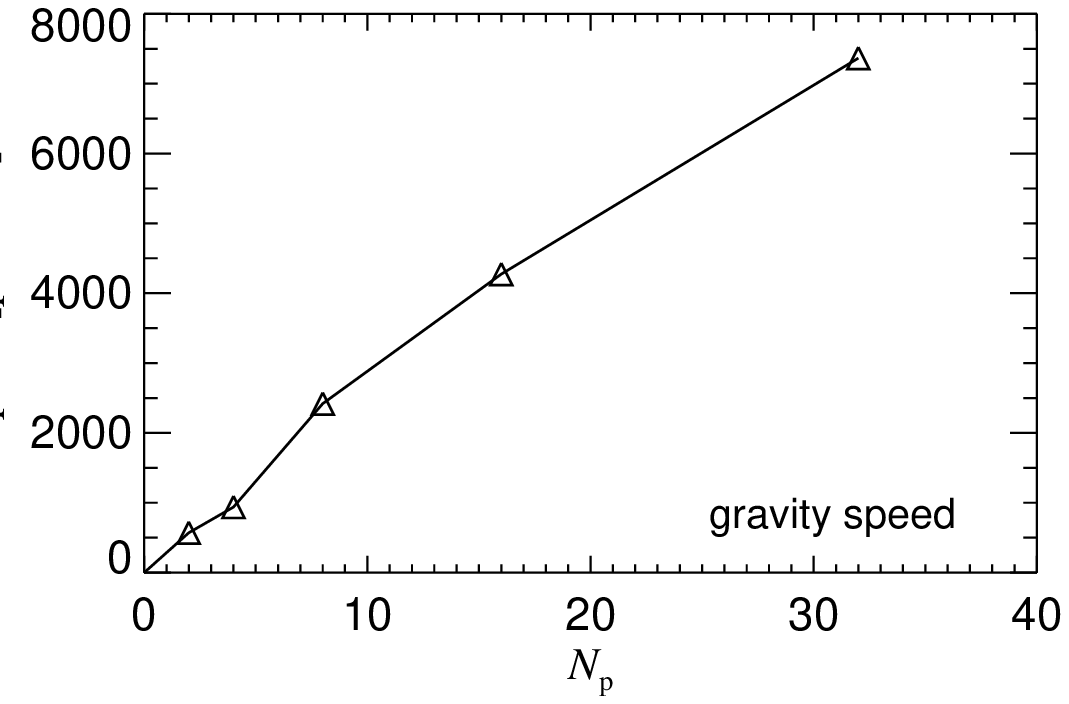}}\\
\resizebox{8.0cm}{!}{\includegraphics{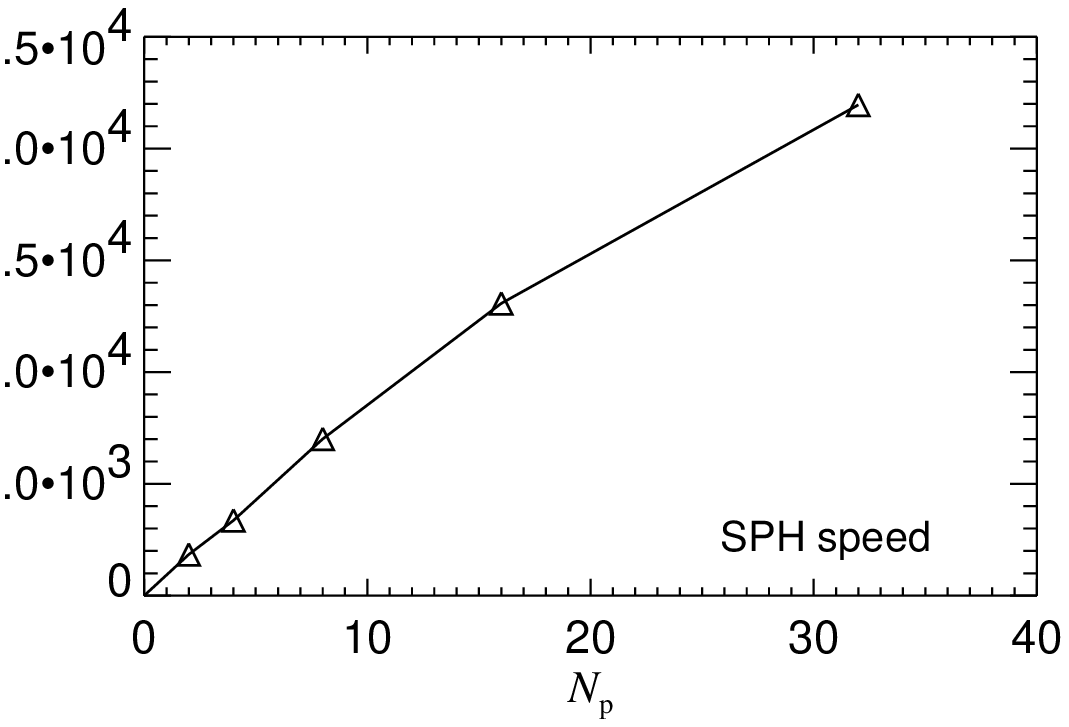}}
\resizebox{8.0cm}{!}{\includegraphics{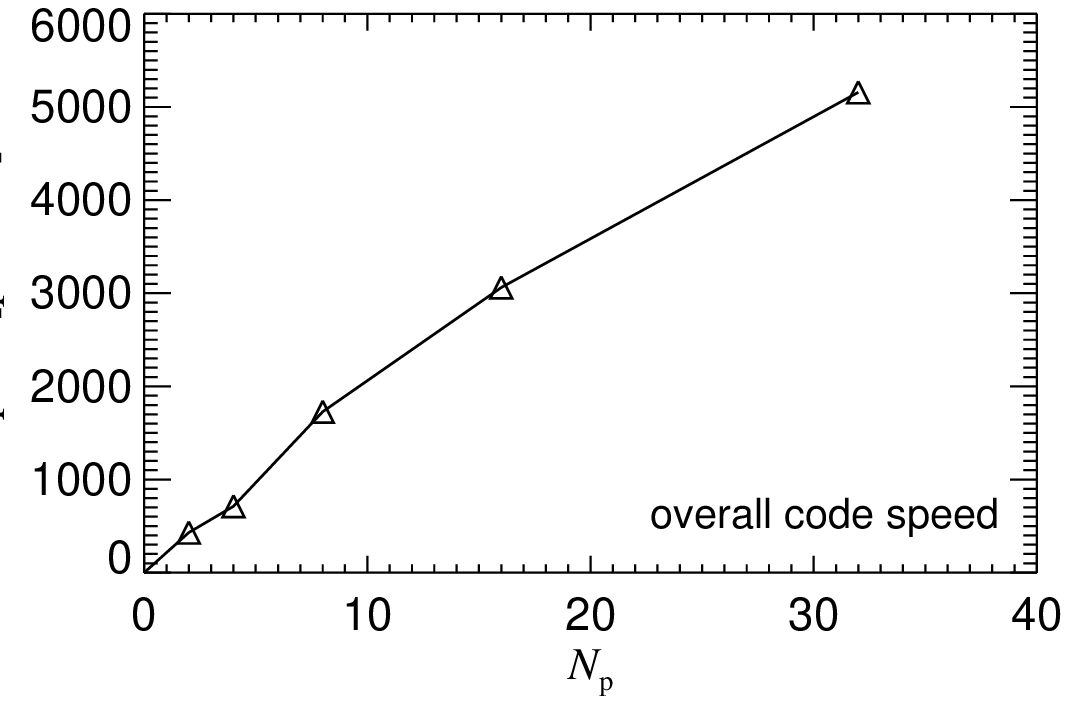}}
\caption{\label{sphspeed} Speed of the code in a gasdynamical
simulation of a periodic $\Lambda$CDM universe, run from $z=10$ to
$z=1$, as a function of the number of T3E processors employed.  The
top panel shows the speed of the gravitational force computation
(including tree construction times). We define the speed here as the
number of force computations per elapsed wall-clock time. The middle
panel gives the speed in the computation of hydrodynamical forces,
while the bottom panel shows the resulting overall code speed in terms
of particles advanced by one timestep per second. This effective code
speed includes all other code overhead, which is less than 5\% of the
total cpu time in all runs.}  
\ec
\end{figure}

We have evolved these initial conditions from $z=10$ to $z=1$, using
2, 4, 8, 16, and 32 processors on the Cray T3E in Garching.  The final
particle distributions of all these runs are in excellent agreement
with each other.

In the bottom panel of Figure~\ref{sphspeed}, we show the code speed
as a function of the number of processors employed. We here define the
speed as the total number of force computations divided by the total
elapsed wall-clock time during this test. Note that the scaling of the
code is almost perfectly linear in this example, even better than for
the set-up used in the previous section. In fact, this is largely
caused by the simpler geometry of the periodic box as compared to the
spherical volume used earlier.  The very absence of such boundary
effects makes the periodic box easier to domain-decompose, and to
work-balance.

The top and middle panels of Figure~\ref{sphspeed} show the speed of
the gravitational force computation and that of the SPH part
separately.  What we find encouraging is that the SPH algorithm scales
really very well, which is promising for future large-scale
applications.

It is interesting that in this test (where $N_{\rm s}=40$ SPH
neighbours have been used) the hydrodynamical part actually consumes
only $\sim 25\%$ of the overall CPU time.  Partly this is due to the
slower gravitational speed in this test compared to the results shown
in Figure~\ref{fig1}, which in turn is caused by the Ewald summation
needed to treat the periodic boundary conditions, and by longer
interaction lists in the present test (we here used our new opening
criterion). Also note that only half of the particles in this
simulation are SPH particles.

We remark that the gravitational force computation will usually be
more expensive at higher redshift than at lower redshift, while the
SPH part does not have such a dependence. The fraction of time
consumed by the SPH part thus tends to increase when the material
becomes more clustered. In dissipative simulations one will typically
form clumps of cold gas with very high density - objects that we think
will form stars and turn into galaxies. Such cold knots of gas can
slow down the computation substantially, because they require small
hydrodynamical timesteps, and if a lower spatial resolution cut-off
for SPH is imposed, the hydrodynamical smoothing may start to involve
more neighbours than~$N_{\rm s}$.

\section{Discussion} \label{secdis}

We have presented the numerical algorithms of our code \gadget,
designed as a flexible tool to study a wide range of problems in
cosmology. Typical applications of \gadget\ can include interacting
and colliding galaxies, star formation and feedback in the
interstellar medium, formation of clusters of galaxies, or the
formation of large-scale structure in the universe.

In fact, \gadget\ has already been used successfully in all of these
areas.  Using our code, \citet{Sp98} have studied the formation of
tidal tails in colliding galaxies, and \citet{Spr99} has modeled star
formation and feedback in isolated and colliding gas-rich spirals.
For these simulations, the serial version of the code was employed,
both with and without support by the \grape\ special-purpose hardware.

The parallel version of \gadget\ has been used to compute
high-resolution N-body simulations of clusters of galaxies
\citep{Spr99c,Spr99b,Yos2000,Yos2000b}. In the largest simulation of
this kind, 69 million particles have been employed, with 20 million of
them ending up in the virialized region of a single object. The
particle mass in the high-resolution zone was just $\sim 10^{-10}$ of
the total simulated mass, and the gravitational softening length was
$0.7\lu$ in a simulation volume of diameter $140\mlu$, translating to
an impressive spatial dynamic range of $2\times 10^5$ in three
dimensions.

We have also successfully employed \gadget\ for two
`constrained-realization' (CR) simulations of the Local Universe
(Mathis~et~al. 2000, in preparation). In these simulations, the
observed density field as seen by IRAS galaxies has been used to
constrain the phases of the waves of the initial fluctuation spectrum.
For each of the two CR simulations, we employed $\sim 75$ million
particles in total, with 53 million high-resolution particles of mass
$3.6\times 10^{9} h^{-1}\msun$ ($\Lambda$CDM) or $1.2\times
10^{10}h^{-1}\msun$ ($\tau$CDM) in the low-density and
critical-density models, respectively.

The main technical features of \gadget\ are as follows. Gravitational
forces are computed with a Barnes~\&~Hut oct-tree, using multipole
expansions up to quadrupole order. Periodic boundary conditions can
optionally be used and are implemented by means of Ewald summation.
The cell-opening criterion may be chosen either as the standard
BH-criterion, or a new criterion which we have shown to be
computationally more efficient and better suited to cosmological
simulations starting at high redshift. As an alternative to the
tree-algorithm, the serial code can use the special-purpose hardware
\grape\, both to compute gravitational forces and for the search for
SPH neighbours.

In our SPH implementation, the number of smoothing neighbors is kept
exactly constant in the serial code, and is allowed to fluctuate in a
small band in the parallel code. Force symmetry is achieved by using
the kernel averaging technique, and a suitable neighbour searching
algorithm is used to guarantee that all interacting pairs of SPH
particles are always found. We use a shear-reduced artificial
viscosity that has emerged as a good parameterization in recent
systematic studies that compared several alternative formulations
\citep{Th98,Lo98}.

Parallelization of the code for massively parallel supercomputers is
achieved in an explicit message passing approach, using the MPI
standard communication library. The simulation volume is spatially
split using a recursive orthogonal bisection, and each of the
resulting domains is mapped onto one processor. Dynamic work-load
balancing is achieved by measuring the computational expense incurred
by each particle, and balancing the sum of these weights in the domain
decomposition.

The code allows fully adaptive, individual particle timesteps, both
for collisionless particles and for SPH particles.  The speed-up
obtained by the use of individual timesteps depends on the dynamic
range of the time scales present in the problem, and on the relative
population of these time scales with particles.  For a collisionless
cosmological simulation with a gravitational softening length larger
than $\sim 30\lu$ the overall saving is typically a factor of
$3-5$. However, if smaller softening lengths are desired, the use of
individual particle timesteps results in larger savings. In the
hydrodynamical part, the savings can be still larger, especially if
dissipative physics is included. In this case, adaptive timesteps may
be required to make a simulation feasible to begin with. \gadget\ can
be used to run simulations both in physical and in comoving
coordinates.  The latter is used for cosmological simulations
only. Here, the code employs an integration scheme that can deal with
arbitrary cosmological background models, and which is exact in linear
theory, i.e.~the linear regime can be traversed with maximum
efficiency.

\gadget\ is an intrinsically Lagrangian code. Both the gravity and the
hydrodynamical parts impose no restriction on the geometry of the
problem, nor any hard limit on the allowable dynamic range.  Current
and future simulations of structure formation that aim to resolve
galaxies in their correct cosmological setting will have to resolve
length scales of size $0.1-1\lu$ in volumes of size $\sim 100 \Mlu$.
This range of scales is accompanied by a similarly large dynamic range
in mass and time scales.  Our new code is essentially free to adapt to
these scales naturally, and it invests computational work only where
it is needed. It is therefore a tool that should be well suited to
work on these problems.

Since \gadget\ is written in standard ANSI-C, and the parallelization
for massively parallel supercomputers is achieved with the standard
MPI library, the code runs on a large variety of platforms, without
requiring any change. Having eliminated the dependence on proprietary
compiler software and operating systems we hope that the code will
remain usable for the foreseeable future. We release the parallel and
the serial version of \gadget\ publically in the hope that they will
be useful for others as a scientific tool and as a basis for further
numerical developments.

\section*{Acknowledgements}
We are grateful to Barbara Lanzoni, Bepi Tormen, and Simone Marri for
their patience in working with earlier versions of \gadget.  We thank
Lars Hernquist, Martin White, Charles Coldwell, Jasjeet Bagla and
Matthias Steinmetz for many helpful discussions on various algorithmic
and numerical issues. We also thank Romeel Dav\'{e} for making some of
his test particle configurations available to us.  We are indebted to
the Rechenzentrum of the Max-Planck-Society in Garching for providing
excellent support for their T3E, on which a large part of the
computations of this work have been carried out. We want to thank the
referee, Junichiro Makino, for very insightful comments on the
manuscript.

\appendix

\section*{Appendix: Softened tree nodes}

The smoothing kernel we use for SPH calculations is a spline of the
form \citep{Mo85} 
\be W(r;h)=\frac{8}{\pi h^3} \left\{
\begin{array}{ll}
1-6\left(\frac{r}{h}\right)^2 + 6\left(\frac{r}{h}\right)^3, &
0\le\frac{r}{h}\le\frac{1}{2} ,\\
2\left(1-\frac{r}{h}\right)^3, & \frac{1}{2}<\frac{r}{h}\le 1 ,\\
0 , & \frac{r}{h}>1 .
\end{array}
\right.  \ee Note that we define the smoothing kernel on the interval
$[0,h]$ and not on $[0,2h]$ as it is frequently done in other SPH
calculations.

\begin{figure}
\bc
\resizebox{8cm}{!}
{\includegraphics{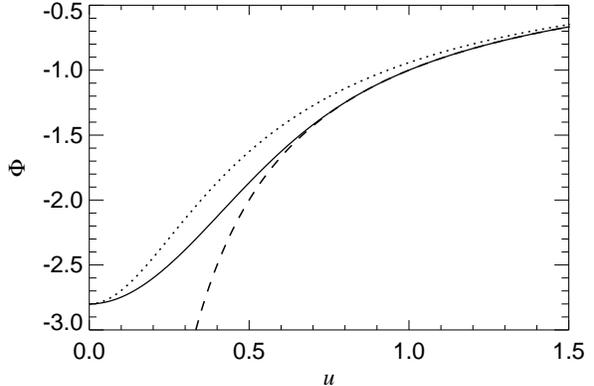}}\\
\caption{Comparison of spline-softened (solid) and Plummer-softened
(dotted) potential of a point mass with the Newtonian potential
(dashed). Here $h=1.0$, and $\epsilon=h/2.8$.
\label{softpot}}
\ec
\end{figure}
\begin{figure}
\bc
\resizebox{8cm}{!}
{\includegraphics{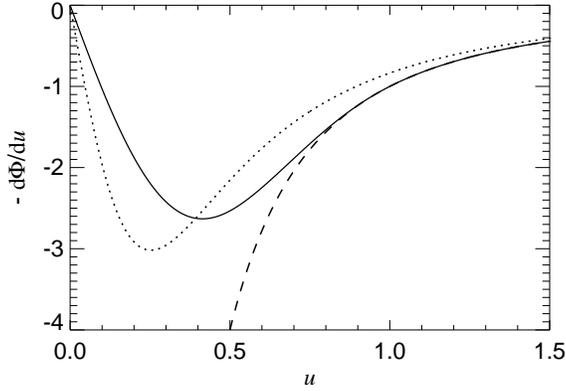}}\\
\caption{Comparison of spline-softened (solid) and Plummer-softened
(dotted) force law with Newton's law (dashed). Here $h=1.0$, and
$\epsilon=h/2.8$.
\label{softforce}}
\ec
\end{figure}

We derive the spline-softened gravitational force from this kernel
by taking the force from a point mass $m$ to be the one resulting from a
density
distribution
$\rho(\vec{r})=m W(\vec{r};h)$. This leads to a potential
\be
\Phi(\vec{r})
=G   \frac{m}{h} W_2\left(\frac{r}{h}\right)
\ee
with a kernel
\be
W_2(u)=
\left\{
\begin{array}{ll}
\frac{16}{3}u^2 - \frac{48}{5} u^4 +\frac{32}{5} u^5 -\frac{14}{5}  ,
&\mbox{$0\le u<\frac{1}{2}$}, \vspace{0.1cm}\\
\frac{1}{15u}+\frac{32}{3}u^2- 16 u^3 +\frac{48}{5}u^4 \\
\mbox{\hspace{1cm}}
-\frac{32}{15}u^5 -\frac{16}{5} ,
& \mbox{$\frac{1}{2}\le u <1$}, \vspace{0.1cm}\\
-\frac{1}{u}, & \mbox{$u \ge 1$}. \\
\end{array}
\right.
\ee
The multipole expansion of a group of particles is discussed in Section
\ref{sectree}.
It results in a potential and force given by equations
(\ref{eqq1}) 
and (\ref{eqqF1}),
respectively.
The functions appearing in equation (\ref{eqqF1}) are defined as
\be
g_1(y) =  \frac{g'(y)}{y} ,
\ee
\be
g_2(y) =  \frac{g''(y)}{y^2} -  \frac{g'(y)}{y^3} ,
\ee
\be
g_3(y) =  \frac{g_2'(y)}{y} ,
\ee
\be
g_4(y) =  \frac{g_1'(y)}{y} .
\ee
Writing $u=y/h$,
the explicit forms of these functions are
\be
g_1(y) = \frac{1}{h^3} 
\left\{ 
\begin{array}{ll}
-\frac{32}{3}+\frac{192}{5}u^2-32u^3, &
\mbox{$u\le\frac{1}{2}$} ,\vspace{0.1cm}\\
\frac{1}{15u^3} - \frac{64}{3} + 48 u \\
\mbox{\hspace{1cm}} - \frac{192}{5}u^2 + \frac{32}{3}u^3, &
\mbox{$\frac{1}{2}<u<1$},\vspace{0.1cm}\\
-\frac{1}{u^3} , & \mbox{$u>1$,}
\end{array}
\right.
\ee
\be
g_2(y) = \frac{1}{h^5} 
\left\{ 
\begin{array}{ll}
\frac{384}{5}- 96 u, &
\mbox{$u\le\frac{1}{2}$},\vspace{0.1cm}\\
-\frac{384}{5} - \frac{1}{5u^5} +\frac{48}{u} +32 u, &
 \mbox{$\frac{1}{2}<u<1$},\vspace{0.1cm}\\ 
\frac{3}{u^5}, & \mbox{$u>1$} ,
\end{array}
\right.
\ee
\be
g_3(y) = \frac{1}{h^7} 
\left\{ 
\begin{array}{ll}
- \frac{96}{u}, &
\mbox{$u\le\frac{1}{2}$},\vspace{0.1cm}\\
\frac{32}{u} + \frac{1}{u^7} -\frac{48}{u^3},  & \mbox{$\frac{1}{2}<u<1$},\vspace{0.1cm}\\
-\frac{15}{u^7} ,& \mbox{$u>1$},
\end{array}
\right.
\ee
\be
g_4(y) = \frac{1}{h^5} 
\left\{ 
\begin{array}{ll}
- \frac{96}{5}(5u-4) , &
\mbox{$u\le\frac{1}{2}$}, \vspace{0.1cm}\\
\frac{48}{u} - \frac{1}{5u^5} -\frac{384}{5} +32u ,  & \mbox{$\frac{1}{2}<u<1$},\vspace{0.1cm}\\
\frac{3}{u^5}, & \mbox{$u>1$} .
\end{array}
\right.
\ee

In Figures \ref{softpot} and \ref{softforce}, we show the
spline-softened and Plummer-softened force and potential of a point
mass. For a given spline softening length $h$, we define the
`equivalent' Plummer softening length as $\epsilon=h/2.8$. For this
choice, the minimum of the potential at $u=0$ has the same depth.

\setlength{\itemsep}{0pt}
\bibliographystyle{mnras}
\bibliography{paper}

\end{document}